\documentclass[acmsmall]{acmart}

\usepackage{graphicx}
\usepackage{xspace}
\usepackage{amsmath}
\usepackage{graphicx}
\usepackage{booktabs}
\usepackage{subcaption}
\usepackage[most]{tcolorbox}
\usepackage[HTML]{xcolor}
\usepackage[linesnumbered,ruled,vlined]{algorithm2e}
\usepackage[absolute,overlay]{textpos}
\usepackage{enumitem}
\usepackage{hyperref}

\newcommand{\toolname}{\textsc{Muse}\xspace}
\newcommand{\chartist}{\textsc{Chartist}\xspace}

\newcommand{\Sec}[1]{Sec. \ref{#1}}
\newcommand{\Eq}[1]{Eq. \ref{#1}}
\newcommand{\Fig}[1]{Fig. \ref{#1}}

\newcommand{\cmark}{\textcolor{green}{\checkmark}}
\newcommand{\xmark}{\textcolor{red}{\textsf{X}}}

\acmJournal{TIIS}

\setcopyright{none}             
\renewcommand\footnotetextcopyrightpermission[1]{} 

\begin{document}

\title{Integrating Multi-Source Feedback in Computational Design}

\author{Francisco Erivaldo Fernandes Junior}
\email{fernandesfefj@ita.br}
\orcid{0000-0003-2301-8820}
\affiliation{%
  \institution{Instituto Tecnológico de Aeronáutica}
  \country{Brazil}
}

\author{Thomas Langerak}
\orcid{0000-0003-2536-0208}
\affiliation{%
  \institution{Aalto University}
  \country{Finland}}
\email{thomas.langerak@aalto.fi}

\author{Mira Ker{\"a}nen}
\orcid{0009-0008-6234-2694}
\affiliation{%
  \institution{Aalto University}
  \country{Finland}
}
\email{mje.keranen@gmail.com}

\author{Danqing Shi}
\orcid{0000-0002-8105-0944}
\affiliation{%
 \institution{Lund University}
 \country{Sweden}
}
\email{shidanqingnet@gmail.com}

\author{Ardak Alipova}
\orcid{0009-0004-3653-9670}
\affiliation{%
  \institution{Nazarbayev University}
  \country{Kazakhstan}
}
\email{ardak.alipova@nu.edu.kz}

\author{Antti Oulasvirta}
\orcid{0000-0002-2498-7837}
\affiliation{%
  \institution{Aalto University}
  \country{Finland}}
\email{antti.oulasvirta@aalto.fi}

\renewcommand{\shortauthors}{Fernandes Junior, Langerak et al.}

\begin{abstract}
	In real-world design practice, evaluations rarely rely on a single source of judgment. Designers routinely combine expert opinions, empirical studies, and computational models, each with distinct strengths and limitations. While machine learning offers methods to integrate multiple feedback sources, these approaches remain largely inaccessible to designers without technical expertise. In this paper, we explore how to integrate multiple feedback sources, primarily through: (1) a practical approach for multi-source integration, and (2) its implementation in \toolname, a no-code tool that allows designers to combine and balance diverse sources. Our technical findings show that independent modeling of multiple evaluation sources enables exploration across heterogeneous feedback, accommodates different evaluation speeds, surfaces disagreements between sources, and supports an adaptable evaluation setup that designers can reconfigure during their process. In a visualization design study, participants navigated their own judgments alongside simulator feedback, reporting a perception of enhanced confidence and flexibility. Our results highlight the viability of multi-source integration to support computational design, offering a step toward bridging the gap between advanced optimization methods and design practice.
\end{abstract}




\begin{teaserfigure}
    \centering
    \includegraphics[width=\linewidth]{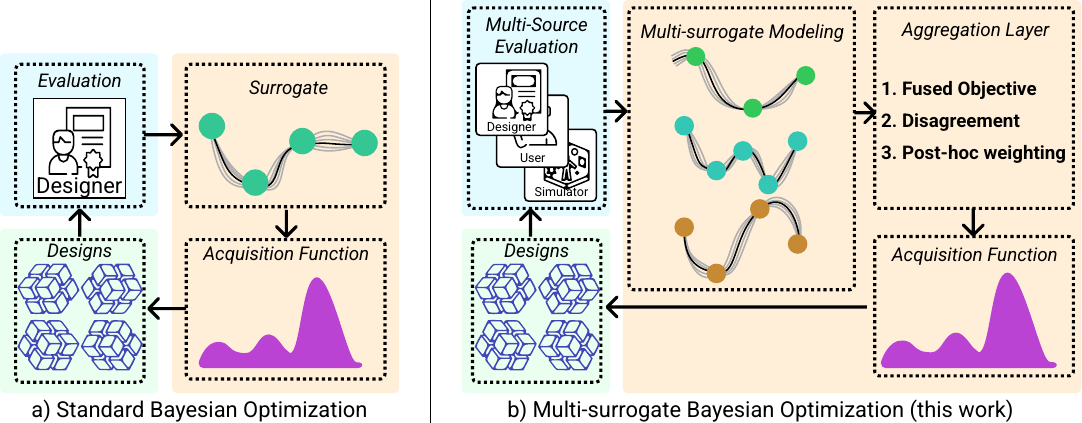}
    \caption{
    We study the integration of multi-source feedback in computational design. a) In standard Bayesian optimization, all evaluative inputs are combined into a single surrogate model, which can obscure differences among sources. b) Using separate source-specific surrogates allows each source’s information to be retained and flexibly integrated, which allows exploration of how heterogeneous feedback can be combined, compared, or weighted, to explicitly control how individual sources contribute to the optimization loop.}
    \label{fig:system-overview}
    \Description{Integration of single- versus multi-source evaluation in design. a) Standard Bayesian optimization treats all evaluative input as a single source, combining them into one surrogate. Differences among sources are lost, making it unclear how each contributes to design guidance. b) Multi-source evaluation retains separate surrogates for each source (e.g., designer, user, simulator), which are then aggregated. This aggregation enables the exploration of how heterogeneous feedback can be combined, compared, or weighted, providing a mechanism to inspect how individual sources contribute to the optimization loop.}
\end{teaserfigure}

\maketitle

\section{Introduction}
\citeauthor{shneiderman2016new}~\cite{shneiderman2016new} argues that the increasingly complex challenges of our time require more integrated approaches to design, where applied and basic sciences converge. In human–computer interaction (HCI), this challenge is especially evident in iterative design, where multiple candidate solutions must be generated and evaluated. In practice, designers rarely rely on a single source of feedback; instead, they perform \emph{triangulation} by consulting multiple sources~\cite{lazar_research_2017,shneiderman2016new}. These sources span analytical methods (e.g., heuristic evaluation~\cite{nielsen_heuristic_1990,benaida_developing_2023}), empirical studies (e.g., usability testing~\cite{hornbaek_introduction_2025}), and computational models (e.g., user simulations~\cite{yamamoto2022photographic, shi2025chartist}). Even in engineering design, where physics models play a central role, designers’ subjective preferences remain important in evaluating solutions~\cite{umetani_pteromys_2014}.

While multi-source optimization methods exist in machine learning~\cite{poloczek2017multi,ghoreishi2019multi,tao2021multi,kandasamy17a}, they remain largely inaccessible to HCI practitioners without technical expertise. In current HCI practice, Bayesian optimization is applied only to single evaluation signals~\cite{chan2022investigating,koyama2022bo,liao2023human,shahriari2015taking}, meaning designers cannot incorporate multiple heterogeneous sources, such as subjective judgments, simulator outputs, or expert opinions, within the optimization process. This creates a practical gap: designers cannot easily combine diverse evaluative signals within existing HCI tools. Understanding how to integrate multiple sources in a way that is transparent, flexible, and usable is therefore critical for supporting informed and reflective design decision-making. Importantly, in HCI, the goal is not simply to find a single ``best'' solution, but to support exploration, reflection, and decision-making across multiple heterogeneous signals.

In this paper, we address this gap by enabling multi-source optimization for practical design tasks. We adapt standard Bayesian optimization to handle multiple heterogeneous evaluation sources in a simple and accessible way, preserving the distinct contribution of each. As illustrated in~\Fig{fig:system-overview}, we maintain an independent surrogate model for each source and combine their predictions into a fused surrogate that guides the optimization loop. We combine predictions through linear aggregation, which offers three key advantages. First, it is principled and avoids arbitrary assumptions about how sources should interact. Second, it is explainable: designers can understand how each source contributes and adjust weights based on their domain knowledge. Third, our empirical results across heterogeneous conditions demonstrate the effectiveness of this assumption. Our contribution is therefore not predictive superiority under idealized conditions, but rather the flexibility to handle heterogeneous sources in scenarios where single-surrogate approaches cannot easily adapt.

For practical integration, we embed this multi-source optimization approach into a no-code design tool, \toolname\footnote{\url{https://github.com/feferna/MUSE}} (Multiple Source Evaluation), which allows designers to work directly with multiple evaluative sources without optimization expertise. \toolname presents designs visually, and predictions and source conflicts numerically, allowing designers to adjust weights or investigate problematic sources without knowing how Gaussian processes and Bayesian optimization work. This integration serves as an interactive prototype that brings multi-source optimization into a familiar workflow, providing the platform to study how designers navigate heterogeneous feedback within a formative task.

Our work provides new insights into how independent modeling of multiple evaluation sources can support designers. Since our approach remains Bayesian optimization–based, it naturally inherits the capabilities to deal with noisy and unreliable sources, even combining them with more reliable ones. Moreover, it accommodates different evaluation speeds, so designers can integrate fast and slow feedback sources without disrupting their workflow. It surfaces disagreements between sources, making conflicts visible and helping designers make informed decisions. Finally, it supports an adaptable evaluation setup, where designers can add, remove, or adjust sources during the design process, facilitating iterative exploration and experimentation. Thus, our work provides the technical foundation for designers to work with different types of evaluations, from subjective judgments to simulator outputs, and offers a mechanism to reflect on their relative importance. We also provide a series of technical evaluations showcasing how our method can handle this diverse set of evaluation sources.

It is important to distinguish our focus from multi-objective Bayesian optimization (MOBO)~\cite{liao_interaction_2023,mo_cooperative_2024}, which treats each input as a separate and often competing objective and seeks trade-offs along a Pareto front. In contrast, we focus on integrating diverse HCI feedback, such as user judgments, expert opinions, and simulations, into a coherent optimization process. Rather than balancing competing objectives, we focus on providing a practical way for designers to combine heterogeneous signals effectively.

Based on these findings, we make three key contributions. First, we demonstrate how Bayesian optimization can support multiple evaluation sources through a multi-surrogate modeling approach. Second, we show how independent modeling of sources provides structural flexibility while maintaining comparable optimization outcomes to single-surrogate approaches. Third, we provide a no-code interface, \toolname, that enables designers to work directly with multiple sources and allows us to study how they navigate heterogeneous feedback in a formative design task. Together, these contributions advance our understanding of how multi-source integration offers the potential to support flexible and reflective workflows in computational design.

\section{Related Work}
Design optimization in HCI has leveraged a variety of evaluation sources, including hand-crafted objectives, human feedback, user simulations, and LLM-based proxies. While effective in isolation, these approaches either rely on a single evaluation signal or treat multiple objectives in rigid MOBO frameworks, which limits flexibility for interactive, human-in-the-loop design. The following subsections review these evaluation sources, MOBO approaches, and existing optimization methods, highlighting gaps our work addresses.

\subsection{Evaluation Sources in Design Optimization}
Many design optimization approaches in HCI rely on hand-crafted, scalar-valued objectives that encode heuristics or performance metrics, such as accuracy and efficiency in keyboard layouts~\cite{feit2021}, spatial balance in layout tools~\cite{oulasvirta2020combinatorial}, or visual clarity and narrative coherence in visualization and video design~\cite{micallef2017towards,shi2021autoclips}. While effective in structured tasks, such objectives capture only part of design intent, often missing subjective or contextual factors, and their tuning requires expert knowledge. In this paper, we, instead, integrate multiple feedback sources, each providing a complementary but noisy signal about design quality.

Human feedback is particularly valuable when evaluations are subjective or hard to formalize. Human-in-the-loop methods avoid predefined objectives by directly querying users, with Bayesian optimization guiding queries to reduce cost~\cite{shahriari2015taking,frazier2018tutorial}. This has enabled interactive systems in domains such as animation~\cite{brochu2010bayesian}, wearables~\cite{kim2017human}, visual layout~\cite{koyama2017sequential,koyama2020sequential}, photography and gesture keyboards~\cite{yamamoto2022photographic}, and font design~\cite{tatsukawa2025fontcraft}. Frameworks like \emph{BO as Assistant}~\cite{koyama2022bo} further embed designers in mixed-initiative workflows. However, most approaches treat human input as the sole evaluation channel, which limits scalability when preferences are unstable or when objective metrics also matter~\cite{wang2023recent}. 

User simulation complements human feedback by providing step-by-step simulations, from which indices of usability can be computed. Early models such as Fitts' Law~\cite{fitts_information_1954} and the Human Processor Model~\cite{card2018psychology} informed interface design, while modern tools like AIM~\cite{oulasvirta2018aalto} and Perceptual Pat~\cite{shin2023perceptual} offer systematic evaluation utilities. Recent advances employ deep learning and reinforcement learning for tasks including menu search~\cite{dayama2020foraging}, visual saliency~\cite{o2014learning,fosco2020predicting}, typing~\cite{shi2024crtypist,shi2025typoist}, pointing~\cite{keurulainen2023amortised}, selection~\cite{li2023modeling,chen2021adaptive}, annotation~\cite{shi2024understanding}, viewing~\cite{jiang2024eyeformer}, and reading~\cite{shi2025chartist}. While these models offer fast and repeatable evaluations, they rely on assumptions that omit subjective or contextual aspects such as intent or aesthetics. We instead treat them as one source among many, combining their strengths with human and LLM-based feedback rather than relying on them alone.

Recently, large language models (LLMs) have also been explored as evaluation sources in design optimization. By leveraging their pretrained knowledge and generative capabilities, LLMs can act as proxies for subjective judgments or domain expertise, providing fast, low-cost feedback in areas such as visualization critique~\cite{duan_uicrit_2024,pan_vis-shepherd_2025}, UI and layout assessment~\cite{chen2025generative,duan_towards_2023}, and design ideation~\cite{shin_integrating_2023,shin_chatbots_2022}. While promising, such approaches inherit biases from training data and can diverge from human intent~\cite{raiaan_review_2024,wang_aligning_2023}, highlighting the need for approaches that combine LLMs with other sources.

In summary, existing work has explored hand-crafted objectives, human feedback, user simulations, and more recently LLM-based proxies as evaluation sources, but typically in isolation. By examining how these sources can be combined in practice, our work explores how designers interact with and integrate heterogeneous feedback to inform design decisions.

\subsection{MOBO in Design Optimization}
Prior work in computational design and human–AI co-creation has applied MOBO to balance competing goals such as performance versus aesthetics or speed versus reliability~\cite{liao_interaction_2023,mo_cooperative_2024}, often building on classical approaches such as ParEGO~\cite{knowles_parego_2006}. These systems typically employ vector-valued surrogate models to represent objectives jointly and guide search toward Pareto-optimal alternatives~\cite{liu_remarks_2018,suzuki_multi-objective_2020}.

While effective in batch-oriented or simulator-driven settings, MOBO poses challenges for interactive design workflows where feedback sources differ in characteristics. Simulators can provide fast, low-noise metrics, whereas human designers contribute slower, subjective, and rationale-based evaluations. Standard implementations jointly model objectives using multi-output Gaussian processes, often with intrinsic coregionalisation models (ICM)~\cite{bonilla_multi-task_2007,swersky_multi-task_2013}. A representative case is \texttt{BoTorch}’s \texttt{MultiTaskGP}~\cite{balandat_botorch_2020}, which shares a latent kernel across tasks but assumes complete and synchronized observations across all objectives.

In contrast, we study the integration of multiple evaluation sources as independent channels, allowing for heterogeneous sampling rates and incomplete observations. Unlike MOBO, which focuses on trade-offs between distinct objectives, our interest lies in developing a technical architecture to combine diverse feedback sources.

\subsection{Multi-Source Surrogate-based Design Optimization}
Prior work on multi-source Bayesian optimization has focused on engineering simulations. \citet{poloczek2017multi} modeled all sources within a single Gaussian process (GP) using source-specific bias terms. \citet{ghoreishi2019multi} built per-source GPs but fused them into a joint model of the objective and constraints. \citet{tao2021multi} proposed a multi-model approach where component surrogates form a structured composite objective optimized with a system knowledge gradient. Related work on multi-fidelity optimization~\cite{kandasamy17a} also models sources of varying cost and accuracy, but considers they are aligned approximations of the same objective.

These methods assume sources are numerical, aligned, and directly comparable, which allows them to be integrated into a unified GP. Our setting in HCI is different: sources can be heterogeneous, asynchronous, subjective, or incomplete. Instead of collapsing sources into a single joint model, we propose to maintain independent surrogates and combine their predictions during function evaluation. This avoids assuming that all sources can be directly compared, allows disagreements between sources to be revealed, and makes it possible to respect heterogeneous sampling schedules. As a result, our approach offers a flexible technical architecture for integrating diverse evaluation channels. For example, one could combine designer feedback, simulations, and LLM proxies using this setup, a direction that, to our knowledge, has not been fully explored in computational design or HCI.

\section{Multi-Surrogate Bayesian Optimization}
\label{sec:technical-details}
Bayesian optimization is widely used in interactive design systems to support exploration and refinement of design alternatives. To study how multiple sources of feedback can be integrated in such settings, we model each source as an independent surrogate within a Bayesian optimization loop. The posterior means of these surrogates are then aggregated into an evaluation function $f(\mathbf{x})$, which guides candidate selection, evaluation, and surrogate updates.

This section defines the optimization problem in the multi-source setting (\Sec{subsec:multi-source-def}), describes how the aggregated evaluation is used for candidate selection (\Sec{subsec:optimization}), explains how disagreement between sources can be quantified (\Sec{subsec:disagreement}), and illustrates the approach with a case combining preference-based and metric-based sources (\Sec{subsec:pref-based}).

\subsection{Problem Formulation: Design Optimization with Multiple Evaluation Sources}
\label{subsec:multi-source-def}
Designers in HCI often rely on triangulation: they combine results from multiple evaluations such as user studies, simulations, or heuristics to reduce individual biases and highlight consistent patterns. 
The quality of a design, however, is ultimately judged by end-users and cannot be measured directly. 
We therefore model evaluation through multiple sources that provide noisy or biased estimates of design quality and combine them into an aggregated evaluation that supports optimization. 

Conceptually, each evaluation source $E_n(\mathbf{x})$, with $n=1,\dots,N$ and design parameterized by $\mathbf{x}\in \mathbb{R}^d$, can be seen as a noisy and biased estimate of overall design quality. 
In practice, we represent each $E_n(\mathbf{x})$ as an independent GP~\cite{rasmussen_gaussian_2008,shahriari2015taking}:
\begin{equation}
	E_n(\mathbf{x}) \sim S_n(\mathbf{x}) = \mathcal{GP}\bigl(\mu_n(\mathbf{x}), k_n(\mathbf{x},\mathbf{x}^\prime)\bigr),
\end{equation}
where $\mu_n(\cdot)$ is the prior mean and $k_n(\cdot,\cdot)$ a kernel function chosen to match the source properties. 
Each surrogate is fit on source-specific data ${(\mathbf{x}_i,E_n(\mathbf{x}_i))}$, yielding a posterior mean $\hat{\mu}_n(\mathbf{x})$ that can later be used in optimization.

To formalize the relationship between sources and design quality, we assume the latter can be described as a weighted combination of criteria such as comfort, performance, or cost:
\begin{equation}
	\label{eq:ref-quality}
	F(\mathbf{x}) := \sum_{C} w_C \, f_C(\mathbf{x}),
\end{equation}
where $f_C(\mathbf{x})$ is the score of design $\mathbf{x}$ on criterion $C$ with weight $w_C$. 
$F(\mathbf{x})$ is an idealized construct: it represents the overall design quality but cannot be observed directly.

In practice, we approximate $F(\mathbf{x})$ with an aggregated evaluation $f(\mathbf{x})$, defined as a function that combines the posterior means of the per-source surrogates:
\begin{equation}
	\label{eq:aggregated-eval}
	f(\mathbf{x}) = \phi\bigl(\hat{\mu}_1(\mathbf{x}), \dots, \hat{\mu}_N(\mathbf{x})\bigr) \approx F(\mathbf{x}),
\end{equation}
where $\phi(\cdot)$ is an aggregation operator. 
In this work, we focus on the linear case,
\begin{equation}
	\label{eq:optim-this-work}
	f(\mathbf{x}) = \sum_{n=1}^N \alpha_n \, \hat{\mu}_n(\mathbf{x}),
\end{equation}
with $\alpha_n$ as user-specified or adaptively tuned weights. 
This aggregated evaluation $f(\mathbf{x})$ provides the signal for the acquisition function and drives the optimization loop.

\begin{algorithm}[t]
	\caption{Multi-Surrogate Bayesian Optimization}
	\label{alg:multi-surrogate}
	\KwIn{Number of sources $N$; aggregator $\phi(\cdot)$; batch size $B$; design space $\mathcal{X}$.}
	\KwOut{Sequence of designs $\{\mathbf{x}_t\} \subset \mathcal{X}$ approximating design quality via the optimization of an aggregated evaluation $f(\mathbf{x})$ (\Eq{eq:aggregated-eval}).}
	
	\textbf{Initialize:} Evaluate $n_0$ initial designs using all $N$ sources to collect $\mathcal{D}_n = \{(\mathbf{x}, S_n(\mathbf{x}))\}$ for $n = 1, \dots, N$\;
	
	\textbf{Fit surrogate models:} \\
	\For{$n \leftarrow 1$ \KwTo $N$}{
		Fit GP surrogate $S_n(\mathbf{x})$ to dataset $\mathcal{D}_n$\;
	}
	
	\textbf{Form aggregated evaluation from posterior means:} \\
	\[
	f(\mathbf{x}) = \phi\bigl(\hat{\mu}_1(\mathbf{x}),\, \dots,\, \hat{\mu}_N(\mathbf{x})\bigr),\text{where } \hat{\mu}_n(\mathbf{x}) = \mathbb{E}[S_n(\mathbf{x}) \mid \mathcal{D}_n]
	\]
	
	\While{stopping criterion not met}{
		\textbf{(1) Select batch:} Optimize acquisition function to sequentially select $B$ ranked candidates for $b = 1, \dots, B$: \\
			\[
			\mathbf{x}^{(b)}_{\text{cand}} = \mathop{\text{arg\,max}}_{\mathbf{x} \in \mathcal{X} \setminus \{\mathbf{x}^{(1)}_{\text{cand}}, \dots, \mathbf{x}^{(b-1)}_{\text{cand}}\}} f(\mathbf{x})
			\]
		
		\textbf{(2) Evaluate candidates:} \\
		\For{each candidate $\mathbf{x}^{(j)}_{\text{cand}}$}{
			\For{$n \leftarrow 1$ \KwTo $N$}{
				Query $S_n(\mathbf{x}^{(j)}_{\text{cand}})$\;
				Store result: $\mathcal{D}_n \leftarrow \mathcal{D}_n \cup \{(\mathbf{x}^{(j)}_{\text{cand}}, S_n(\mathbf{x}^{(j)}_{\text{cand}}))\}$\;
			}
		}
		
		\textbf{(3) Update surrogates:} \\
		\For{$n \leftarrow 1$ \KwTo $N$}{
			Refit $S_n(\mathbf{x})$ using dataset $\mathcal{D}_n$\;
		}
		
		\textbf{(4) Recompute aggregated evaluation:} \\
		\[
		f(\mathbf{x}) = \phi\bigl(\hat{\mu}_1(\mathbf{x}),\, \dots,\, \hat{\mu}_N(\mathbf{x})\bigr),\text{where } \hat{\mu}_n(\mathbf{x}) = \mathbb{E}[S_n(\mathbf{x}) \mid \mathcal{D}_n]
		\]
	}
	
\end{algorithm}

\subsection{Optimizing with Multiple Surrogates}
\label{subsec:optimization}
The optimization aims to find parameters $\mathbf{x}$ that maximize the aggregated evaluation $f(\mathbf{x})$ in a sample-efficient manner. 
We adopt the standard Bayesian optimization loop of alternating between candidate selection, evaluation, and surrogate updates, but extend it to handle multiple sources. 
Each source is modeled by its own GP surrogate (\Sec{subsec:multi-source-def}), and their posterior means are combined to form $f(\mathbf{x})$. 

At each iteration, an acquisition function selects a batch of candidates by maximizing $f(\mathbf{x})$. 
These candidates are then evaluated by the available sources (e.g., simulators, user studies, or preference queries), and the results update the corresponding surrogates independently. 
Because sources often operate at different speeds and costs, updates can occur asynchronously without waiting for all evaluations to complete. 
The loop continues until a stopping criterion is reached, such as a fixed budget or when a designer accepts a solution. 

Algorithm~\ref{alg:multi-surrogate} summarizes our procedure. 
Our implementation uses the Tree-structured Parzen Estimator (TPE) from \texttt{Optuna}~\cite{akiba2019optuna}, but the formulation is compatible with other Bayesian optimization frameworks such as \texttt{BoTorch}~\cite{balandat_botorch_2020}, where aggregation can be expressed using \texttt{ModelListGP}.

\subsection{Computing Disagreement Scores}
\label{subsec:disagreement}
Modeling each evaluation source with an independent GP not only enables aggregation into $f(\mathbf{x})$ but also makes it possible to explicitly measure how much the sources diverge in their predictions. 
This is a key advantage over single-surrogate approaches, where all signals are collapsed into one model and disagreement is hidden. 

We define a disagreement score $D(\mathbf{x})$ at each design point as the variance of the surrogate posterior means:
\begin{equation}
	\label{eq:disagreement}
	D(\mathbf{x}) = \mathrm{Var}\bigl(\hat{\mu}_{1}(\mathbf{x}), \hat{\mu}_{2}(\mathbf{x}), \dots, \hat{\mu}_{N}(\mathbf{x})\bigr).
\end{equation}
High disagreement indicates that the sources provide divergent estimates in that region, for example due to sparse observations, conflicting biases, or incomplete data coverage. Unlike the aggregated evaluation in \Eq{eq:optim-this-work}, which uses designer-specified weights to drive optimization, this disagreement score uses equal weighting and it is not used in the optimization itself. This allows designers to see the raw disagreement between sources before any reweighting is applied, helping them decide whether and how to adjust the source weights in the aggregated evaluation.

In addition to this score, we compute a disagreement uncertainty region, which also uses equal weighting between sources, that reflects how confident the surrogates are in their divergence. 
Given the standard deviations of the surrogates, $\sigma_{n}(\mathbf{x})$, we calculate the average predictive variance:
\begin{equation}
	\bar{V}(\mathbf{x}) = \frac{1}{N} \sum_{n=1}^N \sigma_{n}(\mathbf{x})^2.
\end{equation}
We then define a confidence band around the disagreement score using $\bar{V}(\mathbf{x})$, such that the interval spans from $D(\mathbf{x}) - \bar{V}(\mathbf{x})$ to $D(\mathbf{x}) + \bar{V}(\mathbf{x})$.

For designers, disagreement scores reveal where sources conflict and why aggregation might be unreliable. 
This information can guide decisions such as whether to collect additional human feedback in a region where simulations and automated metrics disagree, or whether to down-weight a source that consistently diverges. 
In other words, beyond providing a signal for optimization, this method also surfaces the structure of disagreement across sources, offering a mechanism intended to provide greater transparency and control when steering the optimization.

\subsection{Illustrative Case: Preference and Metric-based Sources} 
\label{subsec:pref-based}

To make the general setup concrete, we consider a simple illustrative case with two types of sources: a \emph{preference-based} source $g(\mathbf{x})$ and a \emph{metric-based} source $r(\mathbf{x})$. These categories, common in prior work on Bayesian optimization, highlight how relative feedback (preferences) and absolute scores (metrics from simulators or automated measures) can be incorporated in the modeling process.

\paragraph{Preference-based source.}
Preference is a common metric for optimization in HCI (e.g., \cite{yamamoto2022photographic, brochu2010bayesian, kim2017human, koyama2017sequential, koyama2022bo}). 
We model it using the Plackett--Luce model~\cite{plackett_analysis_1975}, which represents the probability of a set of parameters being chosen over other parameters. 

Let $g_i = g(\mathbf{x}_i)$ represent the latent preference score for design $i$. Suppose a designer is presented with a set of options $A$ and chooses a subset $C \subseteq A$, leaving the remaining options $NC = A \setminus C$. The probability that the designer selects the subset $C$ is modeled as the softmax over the summed scores of chosen and non-chosen options:
\begin{equation}
	P(\text{pick } C) =
	\frac{\sum_{i \in C} e^{g_i}}{\sum_{i \in A} e^{g_i}}.
\end{equation}

Fitting the GP to observed user preferences is done in a maximum a \textit{posteriori} (MAP) manner, combining: 
1) the Plackett--Luce \emph{likelihood} term, capturing how $g(\mathbf{x})$ should reproduce observed choice patterns, and 
2) the GP \emph{prior}, enforcing smoothness and generalization across similar designs. 
Since utilities in Plackett--Luce models are identifiable only up to a positive affine transformation, where shifts and rescalings leave the induced choice probabilities unchanged~\cite{marden_analyzing_1995, zhao_learning_2016}, we rescale the fitted latent $g(\mathbf{x})$ to the unit interval $[0,1]$, propagating its variance accordingly. 
This monotone calibration preserves the ordinal content of preferences while enabling a direct and interpretable fusion with other normalized sources.

\paragraph{Metric-based source.}
Besides preferences, design quality can also be assessed through other evaluation sources such as user studies or simulators~\cite{langerak2024marlui, shi2024crtypist}. 
In these settings, an oracle outputs a performance metric $r(\mathbf{x})$, such as success probability or reward. 
We model $r(\mathbf{x})$ with a GP:
\begin{equation}
	r(\mathbf{x}) \;\sim\; \mathcal{GP}\bigl(\mu_r(\mathbf{x}),\,k_r(\mathbf{x},\mathbf{x}^\prime)\bigr),
\end{equation}
fitting it via standard GP regression on $\{(\mathbf{x}_i, r(\mathbf{x}_i))\}$ pairs.

This illustrative case shows how heterogeneous feedback, such as relative judgments and absolute scores, can be expressed within a common surrogate-based framework. 
By normalizing preferences and modeling metrics on the same probabilistic footing, both sources can be aggregated into $f(\mathbf{x})$, making them directly usable in the optimization loop. 
From a modeling standpoint, this demonstrates how fundamentally different forms of feedback can be integrated without requiring them to be reduced to the same representation in advance. This preserves their distinct advantages while still contributing to a unified optimization process.

\section{Technical Evaluation}
\label{sec:technical-evaluation}
To explore how multiple sources of feedback interact in design optimization, we conducted a series of controlled studies. Designers rarely work with clean, single sources. Instead, they combine feedback sources that differ in reliability, sampling rate, and intent. Following prior work in optimization~\cite{tatsukawa2025fontcraft,yamamoto2022photographic,snoek2012practical}, we use controlled ground-truth (GT) functions as proxies for design objectives. This enables us to manipulate specific stressors while retaining a known optimum for validation.

While single- and multi-surrogate approaches can produce similar predictive means under idealized conditions, our studies focus on scenarios where multi-surrogate modeling offers unique architectural flexibility. We investigated six aspects of multi-source feedback: varying sampling rates (\Sec{sec:sampling}), measuring agreement between sources (\Sec{sec:agreement}), adding and removing sources at runtime (\Sec{sec:flexibility}), reweighting sources mid-optimization (\Sec{sec:reweighting}), scaling to larger numbers of sources (\Sec{sec:number-of-sources}), and handling non-stationary objectives (\Sec{sec:nonstationary}). Sinusoidal mixtures were chosen in Studies~1 and~2 because they yield smooth, interpretable objectives and allow us to visualize GP fits directly. Rosenbrock, Three-Hump Camel, Booth and related benchmark functions~\cite{sharma_metaheuristic_2024} were used in Studies~3 to~6 because they better capture optimization performance, with controllable shifts, scaling, or non-stationary optima. Modeling and budget choices were fixed across all studies to ensure comparability. Detailed definitions of all GTs and implementation details are provided in App.~\ref{app:technical-evaluation}.

\subsection{Study 1: Multi-Pace Sampling}
\label{sec:sampling}

\begin{figure}[!t]
	\centering
	\includegraphics[width=0.9\linewidth]{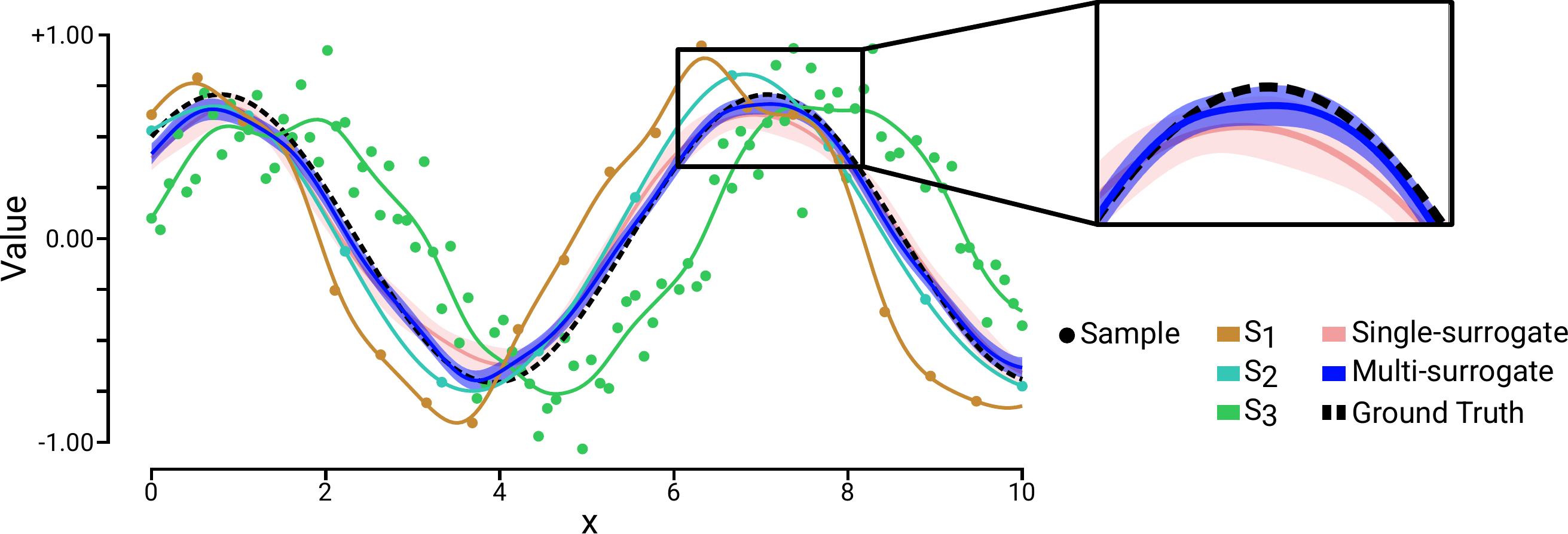}
	\caption{%
		Multi-pace sampling (Study 1): GP posterior fits are shown for a single-surrogate model, which averages all sources and fits a single GP, and a multi-surrogate model, which fits separate GPs for three sources ($S_1$, $S_2$, $S_3$) with different sampling rates and aggregates their predictions. By leveraging dense, noisy sources alongside sparse, high-fidelity ones, the multi-surrogate approach produces more effective design candidates than the single-surrogate baseline.
	}
	\label{fig:multi-pace}
	\Description{This is a line plot comparing single-surrogate and multi-surrogate Gaussian process models in a multi-pace sampling scenario. Three sources with distinct sampling rates, S1 (sparse, high-fidelity), S2 (moderate), and S3 (dense, noisy), are shown with their respective sampled points and fitted curves. The single-surrogate model (pink curve with shaded uncertainty) averages data across sources, while the multi-surrogate model (blue curve with shaded uncertainty) aggregates individually trained surrogates. The ground truth function is represented by a thick black dashed line. An inset zooms into a region where the multi-surrogate model more closely approximates the ground truth, illustrating its superior performance in leveraging heterogeneous data.}
\end{figure}

In their practice, designers may rely on sources with different sampling rates. Oulasvirta et al.~\cite{oulasvirta2020combinatorial} review  heterogenous models (e.g., aesthetic, Fitts’ law models) for combinatorial optimization of layouts. Some are fast, some are slow to run. Feit et al.~\cite{feit2021} and Koyama et al.~\cite{koyama2022bo} rely on user studies and preference feedback, which may be more accurate but necessarily sparse due to cost and participant burden. We evaluate how single- and multi-surrogate methods perform under these conditions.

\paragraph{Setup}
We evaluate a 1D ground truth defined as the average of $\sin(x)$ and $\cos(x)$, 
$F_{\text{true}}(x)=\tfrac{1}{2}(\sin(x)+\cos(x))$, over $x \in [0,10]$. 
Three sources were defined with distinct sampling schedules: $S_1$ (dense, 100 samples), $S_2$ (sparse, 10 samples), and $S_3$ (intermediate, 20 samples). 
Each source was perturbed with Gaussian noise, tested across six regimes: 
$(\sigma_1,\sigma_2,\sigma_3) \in \{(0.05,0.02,0.05),$ $(0.20,0.05,0.10),\ (0.20,0.15,0.10),\ (0.20,0.05,0.25),\ (0.40,0.05,0.10),\ (0.30,0.20,0.25)\}$. 
For each regime, we ran 100 independent seeds, with variance parameters matched in the GP models. 
The single-surrogate baseline was restricted to the slowest sampling rate ($S_2$) by averaging all sources at those points before fitting a single GP. 
In contrast, the multi-surrogate approach maintained a GP per source at its native sampling rate and aggregated predictions by averaging posterior means.

\paragraph{Results}
\Fig{fig:multi-pace} shows one illustrative example of how the multi-surrogate approach more closely approximates the ground truth compared to the single-surrogate baseline. 
Across all six noise regimes and 100 independent seeds per regime, the multi-surrogate method achieved consistently lower mean squared error (MSE) than the baseline in a majority of runs. 
The win rate (percentage of runs where the multi-surrogate achieved lower MSE) was 65\% for $(0.05,0.02,0.05)$, 65\% for $(0.20,0.05,0.10)$, 54\% for $(0.20,0.15,0.10)$, 65\% for $(0.20,0.05,0.25)$, 72\% for $(0.40,0.05,0.10)$, and 57\% for $(0.30,0.20,0.25)$. 
These results confirm that the method offers a consistent advantage across low, moderate, and high noise conditions, though the magnitude of improvement varies. 

\subsection{Study 2: (Dis)Agreement Between Sources}
\label{sec:agreement}

\begin{figure}[!t]
	\centering
	\includegraphics[width=\linewidth]{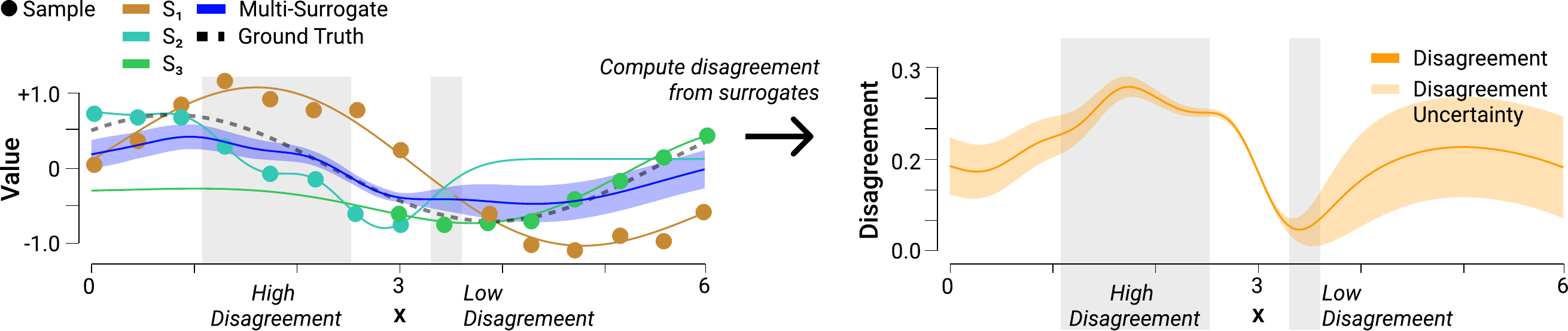}
	\caption{%
		Source (Dis)agreement (Study 2): Unique to multi-surrogate modeling is that we can compute disagreement scores between different feedback sources by comparing their Gaussian processes.
		\textbf{Left:} Individual source data and their GP predictions, plus the averaged multi-surrogate prediction in blue.
		Each source is sampled differently along the domain, creating higher uncertainty in some segments.
		\textbf{Right:} The disagreement (variance across source means). Regions of high variance indicate areas where sources give notably different predictions. The shaded region captures the average predictive variance, allowing us to distinguish disagreement arising from model uncertainty versus genuine differences between the sources.}
	\label{fig:multi-surrogate-disagreement}
	\Description{This is a two-part line plot illustrating source disagreement in a multi-surrogate Gaussian process model. Left: Three data sources (S3, S3, S3) are shown with distinct samples and individual GP regressions. The aggregated multi-surrogate prediction is visualized in blue, closely following the ground truth (dashed black line). Shaded areas highlight regions of higher predictive uncertainty due to sparse or conflicting data. Right: The computed disagreement—defined as the variance across surrogate means—is plotted in orange, with shaded regions indicating disagreement uncertainty. High-disagreement zones correspond to areas where sources diverge, distinguishing modeling noise from true conflict between sources.}
\end{figure}

Different sources might disagree on the quality of a design: the intuition of two designers might differ, or a user study and simulator might show different results \cite{mcdonnell2012accommodating}. It is therefore interesting to quantify where these differences arise, thereby exposing regions in the design space that warrant closer attention.

\paragraph{Setup} 
We again use a sinusoidal ground truth defined as the average of $\sin(x)$ and $\cos(x)$ over $x \in [0,6]$. 
Three sources $S_1$, $S_2$, and $S_3$ were defined with distinct coverage of the domain. 
$S_1$ sampled moderately across the entire range with Gaussian noise $\sigma=0.15$. 
$S_2$ only sampled the left half $[0,3]$ with $\sigma=0.10$, and $S_3$ only sampled the right half $[3,6]$ with $\sigma=0.05$. 
This setting mimics design situations where not all evaluators can assess the full design space, for example when a simulator applies only to certain parameter regions or when user studies are feasible only for selected conditions. 
Following \Eq{eq:disagreement}, predictions were combined by averaging posterior means, and disagreement was quantified as the variance across surrogate means, with average predictive variance used as a baseline to distinguish genuine source conflict from model uncertainty.

\paragraph{Results}
The results of our disagreement study are shown in \Fig{fig:multi-surrogate-disagreement}, which illustrates how disagreement can be visualized alongside each surrogate’s predictions. In the lower subplot, the curve represents disagreement across the domain, highlighting areas where sources provide conflicting estimates. The shaded region indicates average predictive variance, helping distinguish between disagreement due to model uncertainty and genuine differences between sources. We observe that the agreement score reliably captures where surrogates—and by extension, the sources—diverge. By highlighting regions of disagreement, we can help designers focus their data collection on resolving ambiguity and aligning surrogate estimates with the true objective.

\subsection{Study 3: Runtime Adjustment of Sources}
\label{sec:flexibility}

\begin{figure}[!t]
	\centering
	\includegraphics[width=0.9\linewidth]{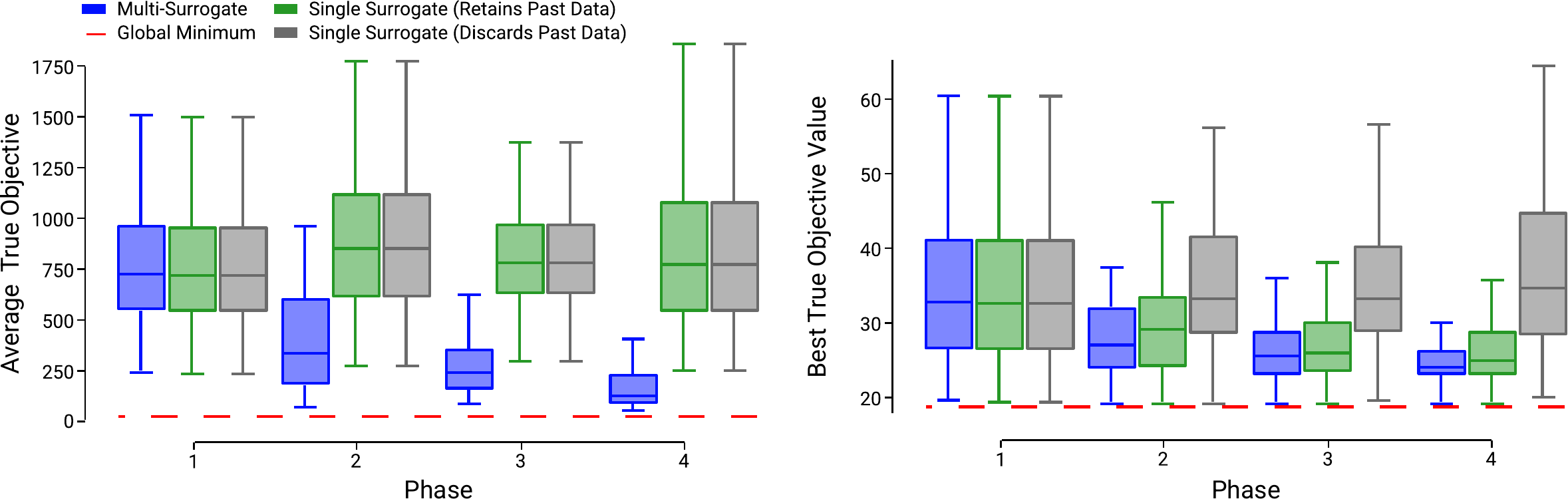}
	\caption{%
		Runtime Source Adjustment (Study 3): Optimization results comparing our multi-surrogate approach against two single-surrogate baselines under dynamic changes in available sources (100 independent runs). Phase 1: $S_1$, Phase 2: $S_1+S_2$, Phase 3: $S_1+S_2+S_3$, Phase 4: $S_1+S_3$.
		\textbf{Left:} Average objective values per phase. The multi-surrogate modeling (blue) outperforms both baselines across all phases. The green baseline restarts optimization but retains past evaluations; the gray baseline restarts and discards all prior data.
		\textbf{Right:} Best objective values found per phase. The dashed line indicates the true global minimum. Multi-surrogate modeling maintains superior performance, especially in Phase 4 when $S_2$ is removed. It reuses all prior knowledge without retraining or discarding information.
	}
	\label{fig:plug_n_play}
	\Description{This figure shows two boxplots comparing the performance of single-surrogate (red) and multi-surrogate (blue) optimization methods across four phases where the combination of data sources changes. Left: Mean objective values show that the multi-surrogate method consistently achieves better average solution quality across all phases. Phase 1 uses only source S1, Phase 2 adds S2, Phase 3 includes all three sources (S1, S2, S3), and Phase 4 removes S2, using only S1 and S3. Right: Best solution values indicate both methods converge toward the global minimum (dashed line), but the multi-surrogate approach yields superior performance when the source set is reduced (Phase 4). Each boxplot summarizes 100 independent optimization runs.}
\end{figure}

During a design project, evaluation sources may be changed intermittently \cite{wright1989evaluation}. By using multi-surrogate modeling, we can maintain this flexibility where traditional single-surrogate methods would require retraining when sources change.

\paragraph{Setup}
We evaluate a four-phase optimization scenario where available sources change dynamically: Phase 1 uses $S_1$ (Rosenbrock), Phase 2 adds $S_2$ (Three-Hump Camel), Phase 3 adds $S_3$ (Booth), and Phase 4 returns to $S_1+S_3$ by removing $S_2$. Each 2D function~\cite{sharma_metaheuristic_2024} is randomly shifted and perturbed with Gaussian noise ($\sigma_1=0.15$, $\sigma_2=0.10$, $\sigma_3=0.05$). Each phase lasts 12 optimization steps (48 total), and experiments are repeated 100 times. We compare against two single-surrogate baselines: Retains Data, which restarts at each phase while keeping prior evaluations, and Discards Data, which fully resets optimizer and history. The multi-surrogate approach instead maintains a surrogate per source, enabling adaptation and reuse when sources change.

\paragraph{Results}
\Fig{fig:plug_n_play} highlights the benefits of multi-surrogate modeling. \Fig{fig:plug_n_play}'s left panel shows it consistently achieves lower average objective values across all phases. In the right panel, while all methods trend toward the global minimum, the performance gap becomes most apparent in Phase 4, when $S_2$ is removed. This method continues to improve by leveraging the surrogate previously trained on $S_2$. The baseline that retains past evaluations performs moderately well but cannot fully exploit prior information due to repeated restarts. The second baseline, which discards all past data, performs the worst, as it loses valuable information at each phase transition and must restart learning from scratch. By modeling sources independently, this approach adapts seamlessly to additions or removals, making it conceptually well-suited for dynamic workflows where single-surrogate methods require ad hoc structural workarounds.

\subsection{Study 4: Mid-Optimization Reweighting}
\label{sec:reweighting}

\begin{figure}[!t]
	\centering
	\includegraphics[width=0.99\linewidth]{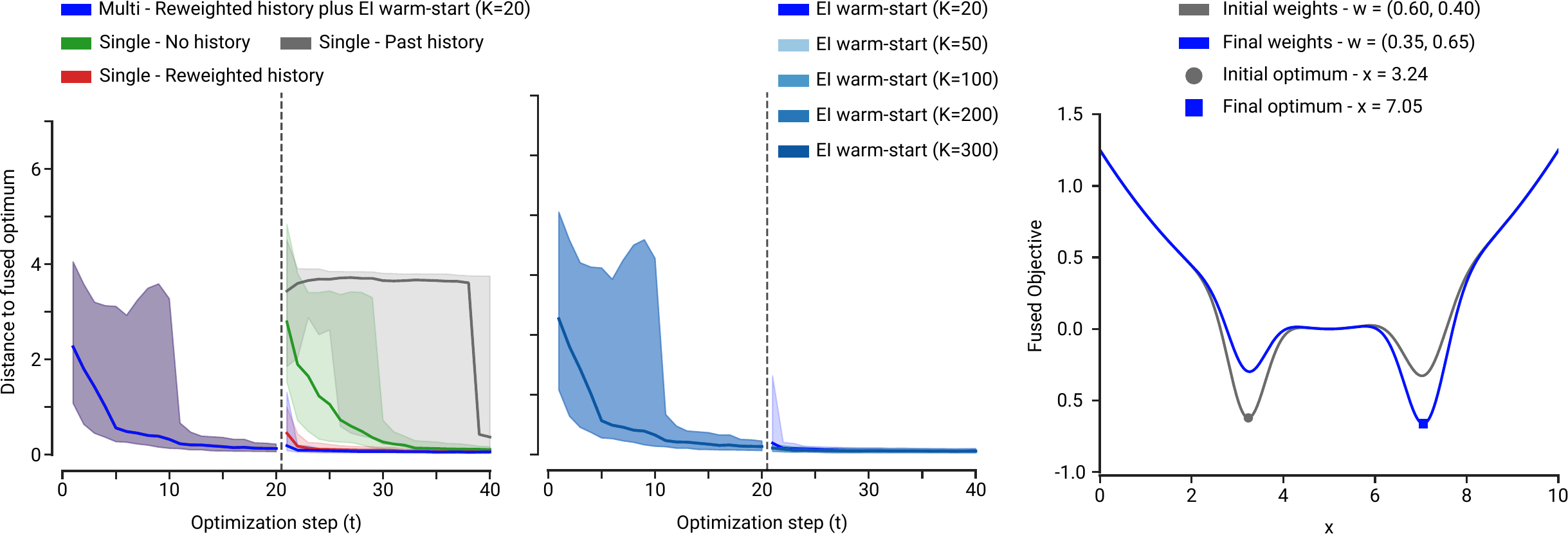}
	\caption{
		Mid-optimization reweighting (Study 4). The dashed line marks the point at \(t{=}21\) when source weights change and the optimizer restarts. \textbf{Left:} comparison of three single-surrogate baselines with the multi-surrogate strategy (EI warm start, $K=20$). \textbf{Center:} effect of varying the number of EI initialization points in the multi-surrogate method. \textbf{Right:} fused source minima before and after reweighting. Shown is the moderate-shift case, which best reflects real design tasks where sources partly agree; results for smaller and larger shifts appear in App.~\ref{app-sec:reweighting}.}
	\Description{This figure consists of three plots illustrating mid-optimization reweighting in Study 4. Left panel: Line plots show optimization step versus distance to the fused optimum. The dashed vertical line marks step 21, where source weights change. The multi-surrogate method (blue) converges quickly, while three single-surrogate baselines (gray, green, red) perform worse, particularly after reweighting. Shaded regions indicate variability. Center panel: Line plots show optimization step versus distance to the fused optimum for different numbers of expected-improvement (EI) warm-start points, ranging from 20 to 300. All converge after the reweighting at step 21, with more warm-start points reducing variance. Right panel: Line plots show fused objective value versus design parameter x. The initial weighting (gray) yields an optimum around x = 3.2, while the reweighted case (blue) shifts the optimum to x = 7.0.}
	\label{fig:reweighting}
\end{figure}

The weight given to different evaluation sources may change as design progresses. For instance, early iterations may rely more on simulations for speed, while later stages emphasize designer judgments. This study evaluates how well multi-surrogate modeling adapts to such mid-process reweighting, compared to single-surrogate baselines.

\paragraph{Setup.}
We perform 100 experiments with Optuna for $T{=}40$ steps, reweighting source contributions at $t{=}21$. Two sources $S_1$ and $S_2$ are distinct smooth functions, combined as $y(x;t)=w_1(t)S_1(x)+w_2(t)S_2(x)$. Here we focus on the case where the weights shift from $(0.60,0.40)$ to $(0.35,0.65)$, which moves the optimum by a moderate distance. This condition, referred to as the medium shift, is most representative of real design workflows where sources usually align to some degree rather than pulling in entirely different directions. At reweighting, we compare one multi-surrogate strategy against three single-surrogate baselines: restarting from scratch, reusing old fused data, and reweighting past data. The multi-surrogate strategy fits independent GPs, reuses and reweights past data, and warm-starts with $K$ points from expected improvement (EI).

\paragraph{Results.}
Figure~\ref{fig:reweighting} shows that all methods behave identically before reweighting. After the change, the multi-surrogate strategy adapts best, starting closer to the new optimum and maintaining a lower error than all baselines. Supplementary material reports results for small and large shifts, which confirm the same trend but with lower and higher variability, respectively.

\subsection{Study 5: Increasing the Number of Sources}
\label{sec:number-of-sources}

\begin{figure}
	\centering
	\includegraphics[width=0.4\linewidth]{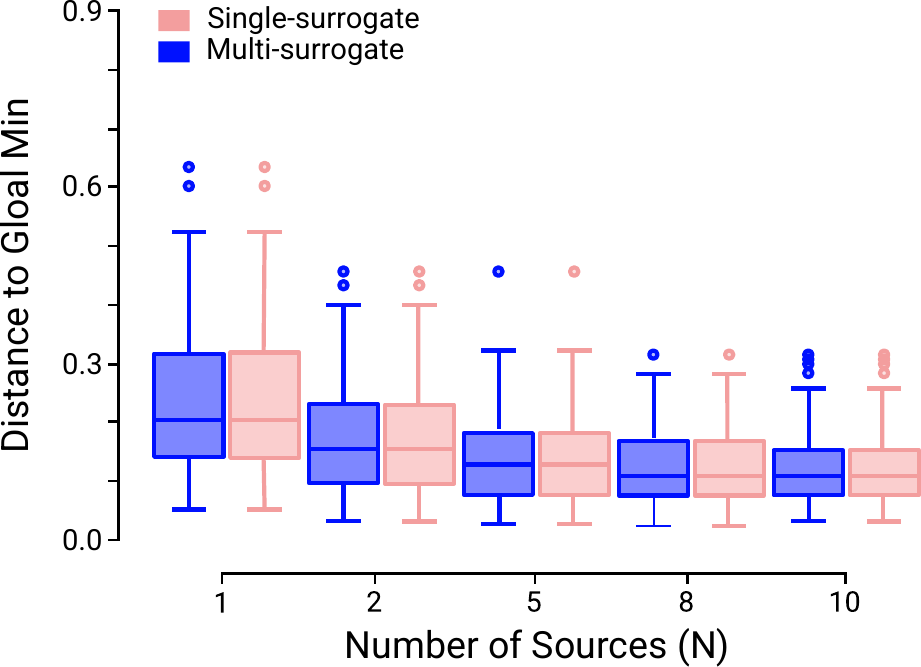}
	\caption{Increasing the number of sources (Study 5): we compare multi- and single-surrogate approaches across $1$, $2$, $5$, $8$, and $10$ evaluation sources, each built from $15$ objective functions. With a fixed seed, both methods produce identical optimization results in terms of Euclidean distance to the global minimum. Increasing the number of sources improves convergence for both. However, the multi-surrogate approach remains more flexible, as it can natively model heterogeneous sources, while the single-surrogate method requires all sources to be merged into a single numeric function.}
	\label{fig:results-combined}
	\Description{A boxplot comparing the performance of single-surrogate (in red) and multi-surrogate (in blue) Bayesian optimization methods across increasing numbers of evaluation sources (1, 2, 5, 8, 10). The y-axis shows the Euclidean distance to the global minimum, with lower values indicating better convergence. For each source count, the multi-surrogate approach produce identical results compared to single-surrogate.}
\end{figure}
Designers often use multiple different sources to evaluate and get feedback on their designs \cite{stone2005user}. We compare the accuracy of single- and multi-surrogate approaches in leveraging multiple evaluation sources to recover the parameters that minimize a ground truth function using Bayesian Optimization.

\paragraph{Setup} We evaluate on the Rosenbrock function, made more realistic by applying random shifts in $[-2,2]$. The true objective is the mean of 15 shifted Rosenbrock functions, while each source consists of the same functions but with randomly assigned weights, making sources distinct from one another and from the ground truth. We test $N={1,2,5,8,10}$ sources under two approaches: the single-surrogate baseline aggregates all outputs into one GP by averaging source values, whereas the multi-surrogate method trains a GP per source and combines predictions into a fused surrogate prediction. Each condition is repeated 100 times, with performance measured as the Euclidean distance between the identified and true minima.

\paragraph{Results} The results are presented in \Fig{fig:results-combined}. Both the multi-surrogate and single-surrogate approaches produce identical optimization trajectories across all cases. Increasing the number of sources from 1 to 10 consistently improves convergence for both methods, as reflected in the reduced Euclidean distance to the true global optimum. This confirms that adding more evaluation sources improves performance, regardless of the surrogate modeling strategy. The multi-surrogate method remains valid even when matching the baseline and, by modeling sources independently, better supports heterogeneous feedback than single-surrogate approaches that force all sources to be scalar.

\subsection{Study 6: Non-Stationary Optimization}
\label{sec:nonstationary}

\begin{figure}[!t]
	\centering
	\includegraphics[width=0.9\linewidth]{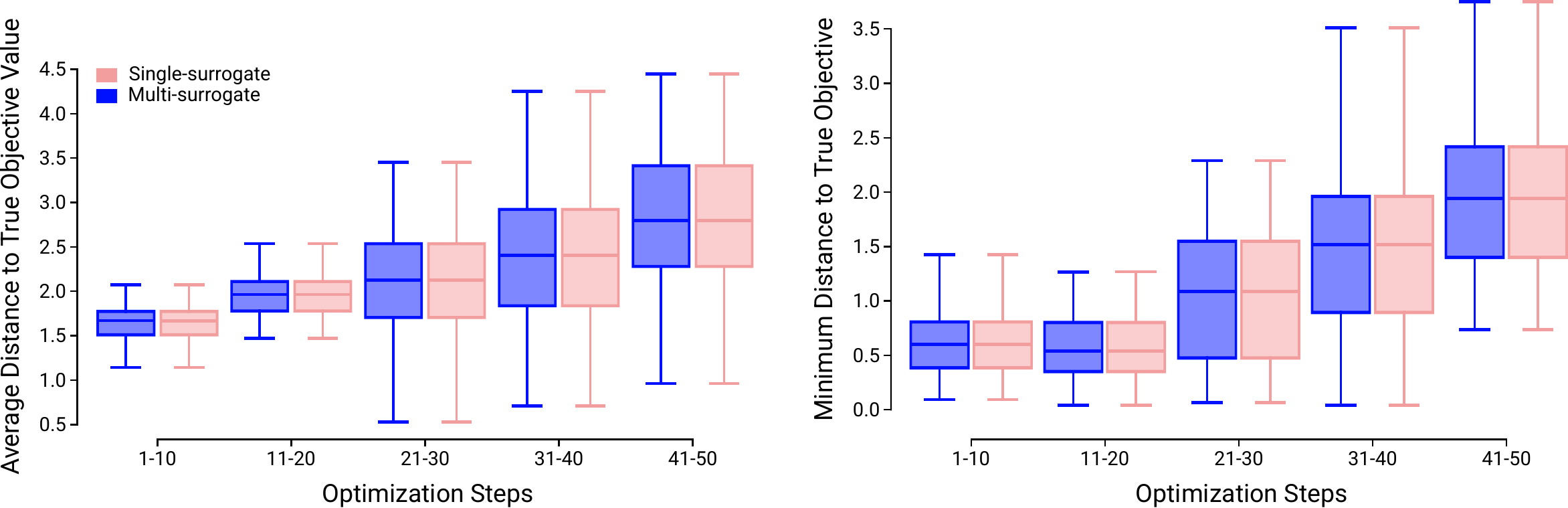}
	\caption{%
		Non-stationary optimization (Study 6): We evaluate the performance of single- and multi-surrogate Bayesian optimization in a setting where the ground-truth optimum shifts every 10 optimization steps. 
		\textbf{Left:} Mean distance to the moving optimum within each block of 10 steps.
		\textbf{Right:} Best (minimum) distance to the moving optimum per block.
		Both methods perform similarly across all blocks, revealing a key limitation: standard Bayesian optimization—regardless of the surrogate model—struggles to adapt in non-stationary settings. This underscores the need for future methods capable of tracking evolving objectives, including in human-in-the-loop design workflows.
	}
	\Description{This figure contains two boxplot panels showing non-stationary optimization results in Study 6. Left panel: Boxplots of average distance to the true objective value over optimization steps, grouped in blocks of 10. Single-surrogate results are shown in red and multi-surrogate results in blue. Both methods start close to the optimum, but distances increase as the optimum shifts every 10 steps. Right panel: Boxplots of minimum distance to the true objective value per block of 10 steps. Red shows single-surrogate and blue shows multi-surrogate. Both methods perform identically, with distance to the optimum increasing over later blocks. The plots illustrate that standard Bayesian optimization, whether single- or multi-surrogate, struggles in non-stationary settings.}
	\label{fig:nonstationary}
\end{figure}

Many real-world design problems are inherently non-stationary: they involve objectives that may shift over  time~\cite{shneiderman2007creativity}. We evaluate how well single- and multi-surrogate optimization methods can adapt to such changing environments.

\paragraph{Setup}
We define a non-stationary ground truth $F_\text{true}(x,t)=\text{Rosenbrock}(x-s(t))$, where the shift $s(t)$ changes every 10 optimization steps. Each source $S_i$ provides a biased variant by applying a fixed, randomly generated offset $b_i$, i.e., $S_i(x,t)=\text{Rosenbrock}(x-s(t)-b_i)$. We compare single- and multi-surrogate approaches with $N=3$ sources over a total budget of 50 optimization steps, repeated for 100 runs. Both methods share the same random seeds, ensuring identical initial points and biases. Performance is measured per block as the mean and best (minimum) Euclidean distance between evaluated solutions and the current optimum.

\paragraph{Results}
As shown in \Fig{fig:nonstationary}, both the single- and multi-surrogate approaches yield the same performance across all five blocks, in terms of both mean (left) and best (right) distance to the moving optimum. Both struggle to consistently track the shifting global minimum introduced every 10 steps. This result highlights a fundamental limitation of standard Bayesian optimization techniques when applied to non-stationary objective functions. Neither single- nor multi-surrogate modeling addresses non-stationary objectives, underscoring the need for new methods and revealing similar limitations in human-in-the-loop design workflows.

\subsection{Summary of Results}
\label{sec:summary}

Table~\ref{tab:comparison-matrix} summarizes the outcomes of all six studies. 
The multi-surrogate approach achieved clear advantages in four cases (Studies~1--4) and matched the baseline in two cases (Studies~5--6). 
This shows that multi-surrogate modeling improves optimization when sources are heterogeneous and/or dynamic, while maintaining comparable performance under idealized or aligned conditions.

\paragraph{Robustness checks.}
We also verified that results hold under different random seeds and noise regimes. 
In Study~1 we repeated all conditions with 100 seeds and multiple noise settings for the three sources, finding that the multi-surrogate consistently reduced error relative to the single-surrogate baseline, with win rates ranging from $54$–$72\%$. 
In Studies~3--6, each configuration was repeated 100 times with different seeds, and Study~4 further examined small, medium, and large weight shifts, confirming that gains were strongest under moderate shifts. 
Together these checks indicate that the observed benefits are not tied to specific seeds or parameter settings.

\begin{table}[t]
	\centering
	\caption{Comparison of single- and multi-surrogate approaches across all technical evaluation studies. \cmark\ indicates the better-performing method; \xmark\ indicates the worse-performing method; “--” indicates equal performance.}
	\label{tab:comparison-matrix}
	\begin{tabular}{lcc}
		\toprule
		\textbf{Study} & \textbf{Single} & \textbf{Multi (Ours)} \\
		\midrule
		1. Multi-Pace Sampling & \xmark & \cmark \\
		2. (Dis)Agreement & \xmark & \cmark \\
		3. Runtime Adjustment of Sources & \xmark & \cmark \\
		4. Mid-Optimization Reweighting & \xmark & \cmark \\
		5. Increasing Number of Sources & -- & -- \\
		6. Non-Stationary Optimization & -- & -- \\
		\bottomrule
	\end{tabular}
\end{table}

\section{\toolname: Multi-Source Evaluation}
\label{sec:walkthrough}

To illustrate how multi-surrogate modeling can be integrated into an interface designed to mirror the familiar flow of Bayesian optimization in HCI, we created \toolname, a no-code proof-of-concept tool for multi-source exploration.
The system integrates simulator-based performance with designer feedback, similar to earlier single-surrogate tools~\cite{chan2022investigating,liao2023human,koyama2017sequential}, but extends them with features unique to the multi-source setting. 
Designers can reweight the sources, adjust optimization parameters on the fly, and use disagreement scores to identify uncertainty or conflicts. 
Candidate designs are evaluated by sources with results aggregated automatically in the backend. 
The workflow runs via a ReactJS frontend and a Python backend using \texttt{FastAPI}.

\begin{figure}[!t] 
	\centering 
	\includegraphics[width=0.8\linewidth]{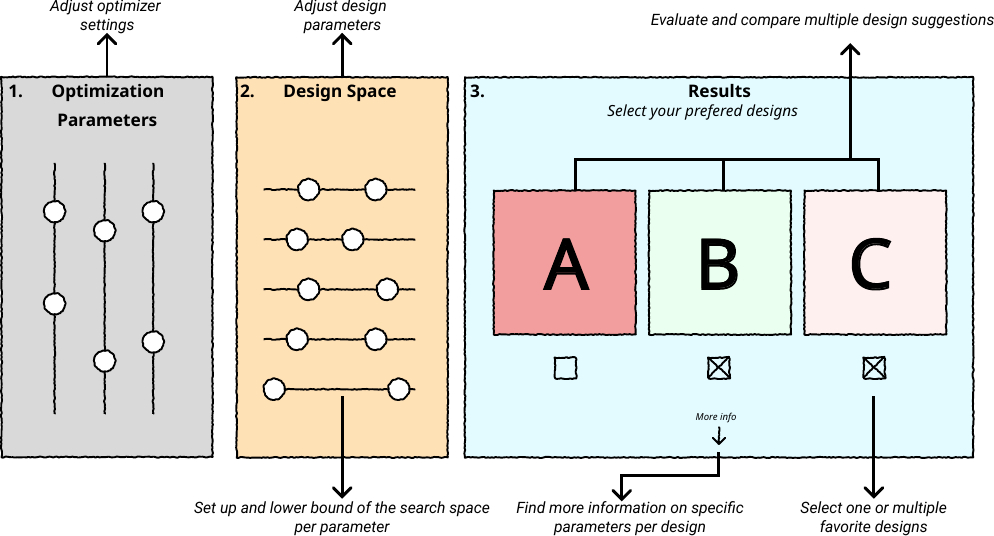}
	\caption{\toolname interface wire-frame for multi-surrogate optimization. (1) Designers configure optimization settings. (2) The design space view allows setting parameter bounds and adjusting design variables in real time. (3) The results panel presents candidate designs together with simulator-provided metrics; designers can inspect details and select one or multiple preferred options. Designs are first evaluated by simulators and then by the designer, with results combined in the backend transparently to the user.}
	\label{fig:ui_walkthrough}
	\Description{This figure shows the Muse interface for multi-surrogate optimization, divided into three panels. Panel 1, labeled Optimization Parameters, let designers configure optimizer settings. Panel 2, labeled Design Space, allows set parameter bounds and adjust design variables. Panel 3, labeled Results, displays candidate designs. Each design can be inspected for more information and selected as one or multiple preferred options. The figure illustrates how designers configure settings, adjust design parameters, and evaluate multiple design suggestions.}
\end{figure}

\begin{figure}[!t]
	\centering
	\includegraphics[width=0.95\linewidth]{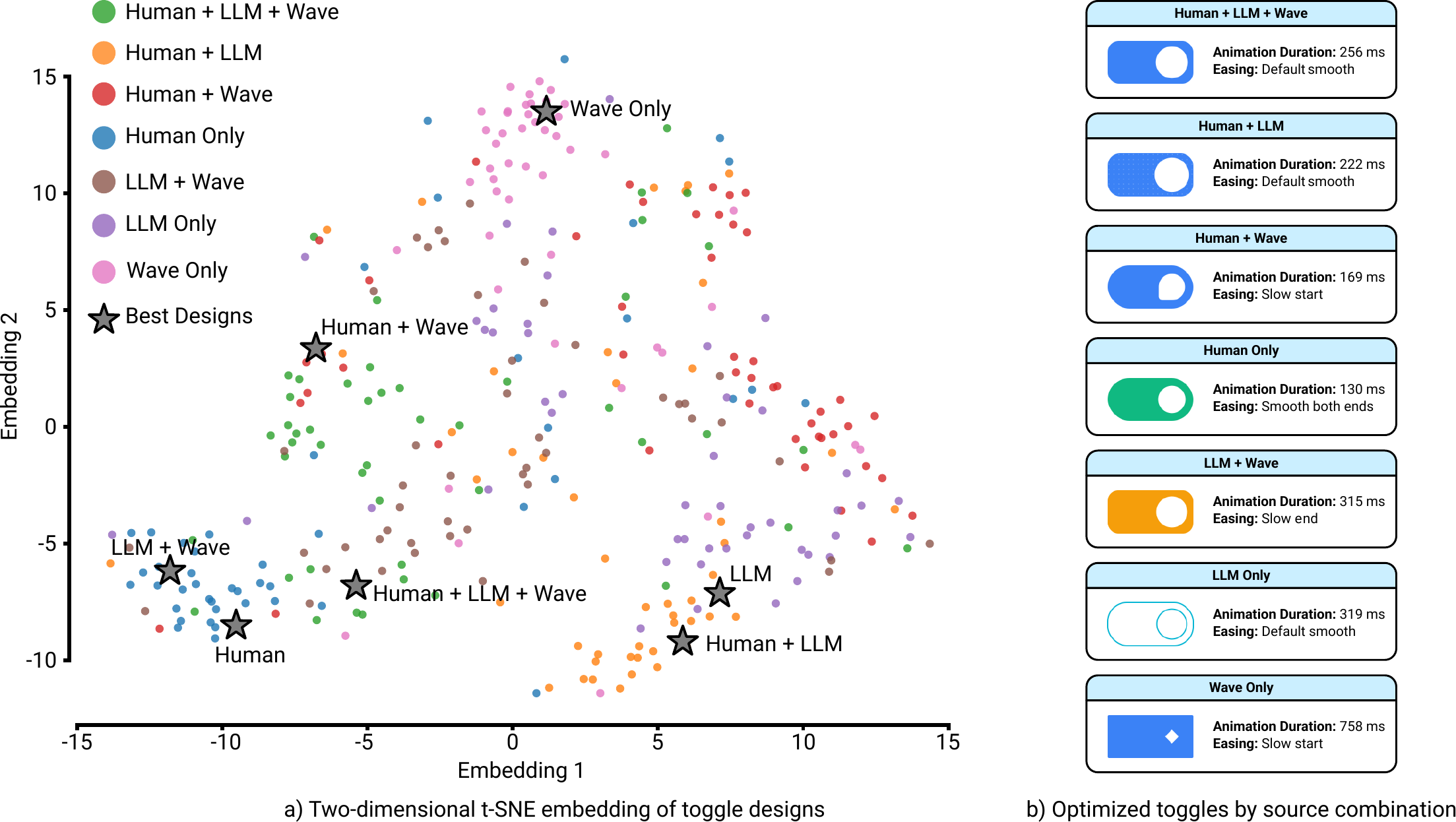}
	\caption{The toggle design demonstration shows how different source combinations shape both the search process and the resulting designs. 
		a) A t-SNE projection of toggle parameters, where each dot represents a candidate evaluated during optimization. Distinct source combinations guide exploration into different regions of the design space. Dark stars mark the best designs identified under each condition. 
		b) The corresponding toggle designs for each condition, illustrating how Human, LLM, WAVE, and their combinations produce different visual and interactive outcomes.}
	\Description{This figure shows results of the toggle design demonstration with two panels. The left panel presents a two-dimensional t-SNE projection of toggle designs, where each dot is a candidate evaluated during optimization. Dots are colored by evaluation source combination: Human, LLM, WAVE, and their pairwise and three-way combinations. Dark stars highlight the best designs found under each condition. The visualization illustrates that different source combinations guide the search into distinct regions of the design space. The right panel displays the final optimized toggle designs for each condition. These examples show how Human, LLM, and WAVE feedback, individually or combined, lead to different visual and interactive outcomes, such as variations in style, color, shape, and animation.}
	\label{fig:toggle-demo}
\end{figure}

\subsection{Walkthrough}
\label{sec:workflow}

\Fig{fig:ui_walkthrough} shows a wire-frame of the interface and workflow of \toolname. Designers first configure the design space, optimization strategy, and evaluation setup by uploading simulators or evaluation scripts. During optimization, simulators and designers evaluate candidate designs, which are modeled with separate GPs and combined into an aggregated evaluation that drives the proposal of new candidates. Disagreement scores highlight uncertainty across sources, and optimization parameters can be updated without restarting. The leftmost panel provides controls for settings such as candidate count or source weights, while the rightmost panel presents candidate designs with visualizations and simulator-derived metrics, allowing designers to select preferred options as feedback. Annotated screenshots of the full interface are provided in App.~\ref{app:walkthrough}.

\subsection{Demonstration: Designing a Toggle with Three Sources}
To illustrate how \toolname can integrate multiple sources, we used it for a toggle design case with three sources of feedback: \emph{Human} (a student with design experience), \emph{LLM}, and a computational aesthetic metric called \emph{WAVE}~\cite{palmer_ecological_2010}. WAVE was chosen because it is a widely used perceptual metric that predicts human judgments of visual aesthetics by modeling how color, symmetry, and balance contribute to perceived quality. 
It provides a quantitative score that can be computed automatically, making it a practical complement to human preference data and LLM-based evaluations.

The toggle’s appearance was defined through seven adjustable dimensions spanning color, style, geometry, and motion. The color palette set the active, inactive, and thumb colors. The visual style controlled the rendering of the track and knob, allowing solid, gradient, outlined, textured, or glass effects, with optional glow or shadow. The thumb shape could vary from circle and square to diamond, hexagon, star, triangle, teardrop, or bean. The roundness of the track corners was determined by the border radius, while the knob size was set as a proportion of the track. Finally, the animation was governed by two parameters: duration and easing, which controlled both speed and smoothness of the transition. Details of the toggle’s design space parameters and the LLM prompts used in the demonstration are provided in the Supplemental Material.

The demonstration focused on optimizing the appearance of Duolingo’s Daily Reminder toggle. 
The human preference emphasized a flat and solid visual style with a greenish background, smooth eased animations, a circular thumb, a relatively large knob proportion, and generous border radii for a rounded look. 
To illustrate how \toolname integrates multiple sources, we performed seven demonstration runs corresponding to the following conditions (shown in Fig.~\ref{fig:toggle-demo}): 
\emph{Human + LLM + WAVE}, \emph{Human + LLM}, \emph{Human + WAVE}, \emph{Human Only}, \emph{LLM + WAVE}, \emph{LLM Only}, and \emph{WAVE Only}. 
Each run was configured for 48 optimization steps with four candidate designs per batch, mirroring a typical setup in standard Bayesian optimization tools but extended to the multi-source case. 

\Fig{fig:toggle-demo}.a shows a two-dimensional t-SNE projection of the toggle parameters sampled across all conditions. The visualization, \Fig{fig:toggle-demo}.b, highlights how different source combinations drive exploration into distinct regions of the design space, while Bayesian optimization is able to identify candidate designs optimized for those specific source profiles. The different source combinations produced distinct outcomes. The \emph{Human Only} run was the only case where the final toggle design matched all of the characteristics sought by the designer, including the flat green style, circular thumb, and large knob with rounded corners. In contrast, the \emph{WAVE Only} run prioritized color palettes over geometry, and since its built-in color preferences diverged from the designer’s green background, it failed to recover the intended look. The \emph{LLM Only} run produced balanced designs but with more variability in style, often selecting shapes and animations consistent with common interface patterns rather than the target preference. Mixed-source runs combined these tendencies: for example, \emph{Human + WAVE} balanced the designer’s preferred geometry with WAVE's color bias, while \emph{Human + LLM + WAVE} broadened exploration and surfaced a set of viable alternatives rather than converging to a single match. Supplementary material also includes additional visualizations (e.g., UMAP projections) that confirm the same patterns of source-specific exploration.

This demonstration illustrates how our integration can flexibly combine human judgment, LLM-based evaluations, and computational metrics to explore different regions of the design space.

\section{Empirical Evaluation with Designers}
\label{sec:evaluation}

The empirical evaluation was designed as a formative study to understand whether designers can work with multi-source feedback through MUSE and how they experience the process. Our goal was not to measure objective design outcomes, but to explore usability, perceived value, and qualitative insights about how designers engage with multiple sources. Specifically, we asked: (i) how do designers work in the multi-source condition, and (ii) does including a computational source change how they experience the design process compared to relying on a single source?

We conducted a user study with designers performing a realistic chart-design task guided by professional-style briefs. This setup offered three key advantages: the use of a realistic design task that reflects professional practice, the recruitment of non-novice designers with prior HCI or design training, and the integration of a cutting-edge computational model alongside human feedback \cite{shi2025chartist}.

\subsection{Method}
\label{sec:designer-eval}
The study used a within-subjects setup to structure the qualitative comparison across two feedback conditions:

\begin{itemize}
	\item \textbf{Single-source (Designer only):} Designers relied solely on their aesthetic judgment and prior experience. The Bayesian optimizer used only designer preferences to suggest subsequent design iterations.
	\item \textbf{Multi-source (Designer+Simulator):} Designers received combined feedback from their own preferences and from \chartist, an attention simulator that predicts how users allocate visual attention when reading charts~\cite{shi2025chartist}. It produces a performance score based on the proportion of simulated fixations falling on task-relevant areas of interest, serving as a computational proxy for chart interpretability.
\end{itemize}

Each participant completed two design tasks (one per condition), with task–condition pairings and order counterbalanced to mitigate potential learning and order effects. This setup allowed us to investigate whether \toolname supports multi-source optimization in a similar way that previous work~\cite{chan2022investigating,liao2023human,koyama2017sequential} supported single-source optimization.

\subsubsection{Participants}
We recruited 15 experienced designers (12 female, 3 male), aged 23–33 years ($25.27\pm3.13$), all of whom had completed at least one course or project in HCI or interaction design. Programming experience was not required. Participants provided informed consent and were compensated with a €16 gift card. The total duration of each study session was approximately one hour.

\subsubsection{Tasks}
Participants used \toolname to complete two tasks guided by a realistic professional design brief (see \Fig{fig:qualitative-results}):
\begin{itemize}
	\item \emph{Theme Park Attendance Task:} Participants designed a visualization clearly depicting worldwide attendance at major theme parks.
	\item  \emph{Paralympic Competitors Task:} Participants created a visualization illustrating the number of Paralympic competitors per country.
\end{itemize}
Both tasks required participants to iteratively refine visual parameters (e.g., bar width, aspect ratio, font sizes, RGB color values) to achieve a clear and professional-quality visualization.

\subsubsection{Procedure}
Each session began with a brief tutorial and warm-up exercise to familiarize participants with the \toolname tool and study procedures. After the warm-up, participants completed two design tasks, each corresponding to a different feedback condition (Designer only or Designer+Simulator). Participants iteratively refined their designs, supported by the optimizer's suggestions, until satisfied with the result. 
Each task was followed by a post-task questionnaire assessing user experience and cognitive workload. The session concluded with a semi-structured interview exploring participants’ overall impressions, satisfaction, and perceived value of the tool.

\subsubsection{Measures}
We collected subjective feedback regarding user experience. We captured participant ratings on a 5-point Likert scale for self-reported satisfaction with both the tool and the resulting designs.

\subsection{Results}
\begin{figure*}
	\centering
	\includegraphics[width=\linewidth]{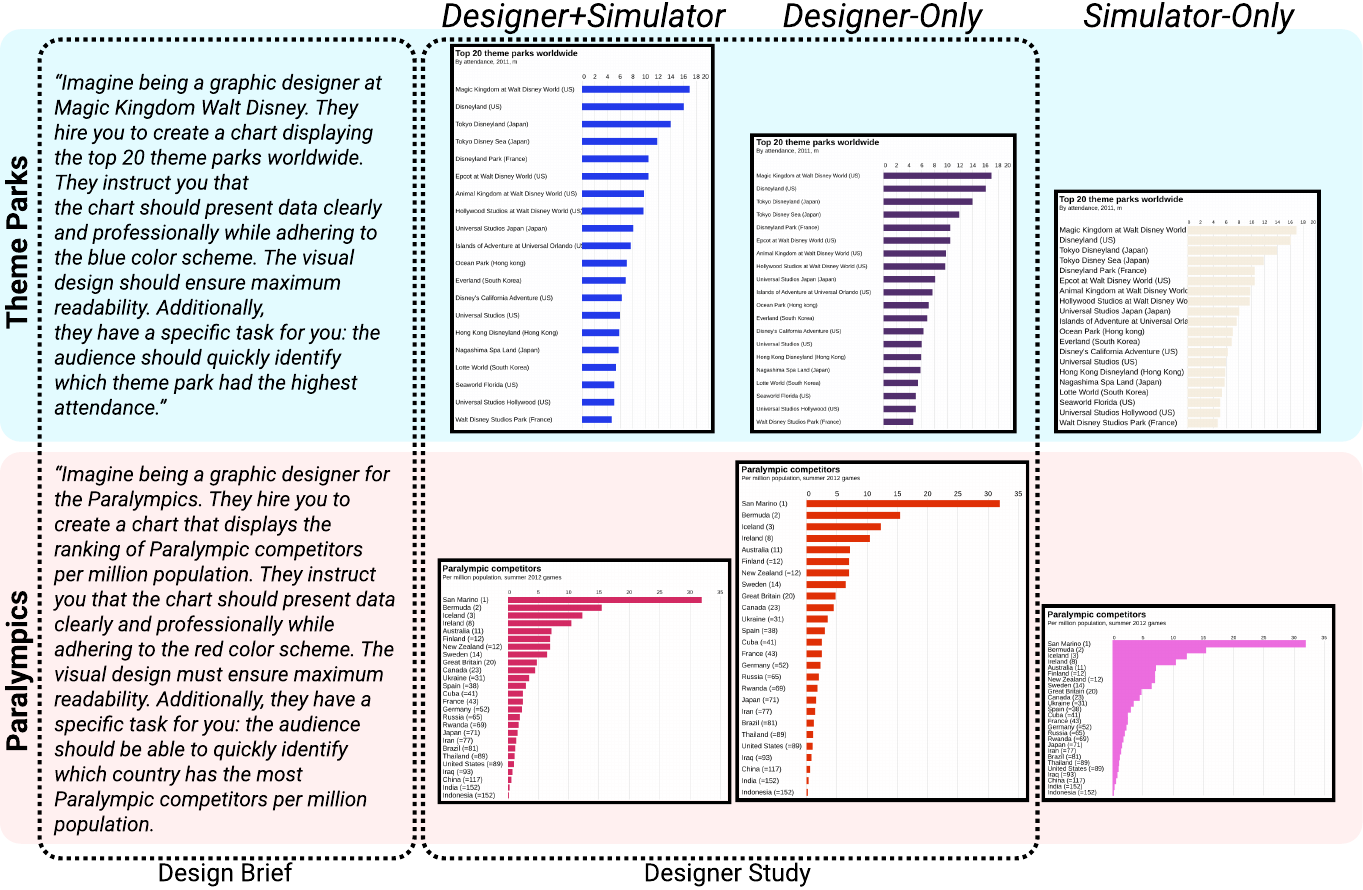}
	\caption{Outcomes of the designer evaluation for different participants, all conditions and tasks. It also includes sample outcomes for a Simulator only baseline.}
	\label{fig:qualitative-results}
	\Description{This is a matrix-style figure showing outcomes from a designer evaluation study across two task domains—Theme Parks and Paralympics—and three conditions: Designer+Simulator, Designer-Only, and Simulator-Only. The left column presents two design briefs: one for creating a theme park attendance chart using a blue color scheme, and one for ranking Paralympic competitors using a red scheme. The central column displays example chart outputs generated by designers with and without simulator feedback. The right column shows outputs generated solely by the simulator. Charts vary in color scheme, label clarity, and data emphasis. The figure highlights differences in visualization quality and task adherence across study conditions.}
\end{figure*}

Some outcome examples can be seen in \Fig{fig:qualitative-results}, all from different participants. 
The figure also includes designs produced in a Simulator-only baseline, where the optimizer relied exclusively on \chartist. 
Because the simulator prioritizes ease of reading bar charts rather than adherence to design briefs, these outputs often ignored the specified color schemes and placed bars very close together, highlighting the difference between computational proxies and human design intent.

\subsubsection{Exploratory Quantitative Analysis}
The following quantitative results are exploratory. Given the formative scope and sample size, they should be interpreted as complementary to the qualitative findings rather than as definitive statistical evidence. A non-parametric ART (Aligned Rank Transform) ANOVA \cite{wobbrock2011aligned} was used to analyze the effects of optimization source (Designer only vs. Designer+Simulator) and task assignment on participant ratings. The analysis showed no effect on the feedback type ($F(1, 103) = 0.61, p = .436$), task ($F(1, 13) = 1.21, p = .291$), or feedback type $\times$ task interaction ($F(1, 103) = 2.54, p = .114$). For completeness we conducted a Wilcoxon post-hoc, which showed no significance or effect sizes for any of the questions (all $p>.4$ and $r<.20$). 

\subsubsection{Tool preference}
During the exit interview we found that participants valued have the simulator in addition to their own preferences. Most participants gave confidence as reason, for example: "I would go with the tool [Designer+Simulator] because I would have an assurance that I will have some assistance from the optimizer" (P4). Similarly, designers noted a perception of flexibility when using the tool:  "It gave you an option to kind of consider both aspects [both data sources], so it had more flexibility" (P7). Finally, a participant mention perceived efficiency: "I think for the first one [Designer+Simulator] I didn't have to do as many rounds, it kind of did what I wanted without me having to actually do it" (P3), however, we did not find any quantitative evidence for this perceived efficiency. 

\subsubsection{System performance}
In contrast, the simulator’s slow performance impacted participants’ design processes: "I didn’t bother doing many iterations; otherwise, I probably would have made more changes and refined it further" (P11). Additionally, many participants did not fully understand how the simulator-based metric functioned. As one noted, they "wished for a little bit more information about what was actually considered in this simulation" (P9), which made it difficult to use the tool effectively during the design process.

\subsubsection{Summary}
The study suggests that the expert designers in our sample were open to engaging with multiple sources during design optimization. At the same time, the results point to a need for transparency: participants expressed a desire to better understand what each source contributes, how its evaluations are generated, and where potential biases may arise, so they can make informed decisions during the process. As a formative study with 15 participants, we do not claim statistical significance. The quantitative results are presented as exploratory findings, complementing the qualitative insights of this study.

\section{Discussion}
\label{sec:discussion}
Multi-surrogate modeling offers clear theoretical and algorithmic advantages over single-surrogate approaches for handling complex multi-source settings. Single-surrogate approaches require all sources to share the same design points and noise levels to match multi-surrogate performance, a condition that rarely holds in real design practice. In contrast, multi-surrogate naturally handles sources with different sampling rates and varying noise levels (Study 1), mismatched domain coverage (Study 2), and dynamic changes to sources (Studies 3 and 4). This setup also supports adaptive weighting of feedback, surfaces disagreements between sources, and allows sources to be added or removed without restarting the optimization process. For example, noisy, frequent sources can be combined with sparse, high-fidelity ones (Study 1), weights can shift smoothly during optimization (Study 4), and regions of conflict between sources highlight areas for further inspection (Study 2). These are precisely the scenarios where designers require workflow flexibility, and our technical evaluations demonstrate that multi-surrogate modeling structurally enables it without sacrificing algorithmic optimization performance.
	
This flexibility stems from our deliberate design choice: linear aggregation with independent surrogates. We do not view this as a limitation but as a design decision validated by our results. Linear aggregation avoids arbitrary assumptions about how sources interact, makes source contributions transparent to designers, and across Studies 1–4 proved sufficient for handling heterogeneous feedback. Non-linear alternatives would introduce opacity, reduce explainability, and offer no demonstrated benefit for the design scenarios we studied. We therefore maintain that linear aggregation is both principled and practical for multi-source optimization in HCI.
	
The same independence that enables flexibility also benefits computational efficiency. Fitting multiple independent surrogates has total time complexity $O(\sum_{n=1}^{N} m_n^3)$, or $O(N \cdot m^3)$ when sources have similar data sizes, compared to $O(m^3)$ for a single surrogate~\cite{rasmussen_gaussian_2008}. However, because the surrogates are independent, they can be fitted in parallel. As a result, the computational time to fit $N$ surrogates is equivalent to fitting a single surrogate. Nevertheless, in practical design contexts, this computational cost is negligible relative to the time required to obtain source feedback. Collecting feedback, whether through user studies, simulations, or physical experiments, often takes minutes, hours, or even days~\cite{langerak2026}. A related concern is whether the approach scales to larger numbers of sources. Study 5 shows that adding sources improves optimization performance up to approximately five sources, after which diminishing returns set in. This diminishing return occurs because additional sources provide increasingly redundant information. Thus, from an algorithmic perspective, multi-source optimization provides meaningful benefits without requiring dozens of distinct sources. Studies 1 and 4 further show that strategic use of partial, incomplete, or reweighted sources matters more than sheer quantity. We therefore consider the flexibility benefits of multi-surrogate modeling to outweigh the modest computational overhead.

The empirical study complements these findings by showing how designers actually engage with multi-source feedback in a realistic design scenario. Participants successfully used \toolname to complete a visualization task while combining their own judgments with feedback from a simulator. They described the additional feedback as providing confidence and flexibility, and appreciated being able to consider “both aspects” of the process. At the same time, they emphasized the need for transparency—understanding how computational evaluations were produced was essential for interpreting and trusting their role in decision-making. These observations align with the motivations behind our study: how to combine multi sources of feedback in a way that respects heterogeneity, exposes conflicts, and supports informed designer choices.

It is also important to note that our multi-surrogate approach and \toolname provide the algorithmic scaffolding to handle unreliable feedback without requiring automated discarding of sources. \toolname keeps designers in control by giving them the ability to adjust source weights based on the disagreement signal and their own judgment. This places control in the hands of designers, who can decide which sources to trust, reduce the influence of those they deem unreliable, or investigate further. Importantly, designers routinely consult different feedback sources at different stages of the design process~\cite{eckert_algorithms_2000, deken_tapping_2012}. Rather than prescribing rigid protocols for when or how to adjust sources, \toolname is designed to allow space for designers to apply their professional judgment. While this combination of surfacing disagreements and enabling manual reweighting was not formally evaluated in our user studies, it represents a critical direction for future work: understanding how designers use these tools to manage unreliable feedback and prevent misleading outcomes in practice.
	
Importantly, \toolname aims to shield designers from the underlying Bayesian optimization mechanics; they interact with source predictions and disagreement measures directly, making decisions based on observable interface outputs rather than mathematical details. The interface presents designs visually, and predictions and sources' conflicts numerically, allowing designers to adjust weights or investigate problematic sources without knowing how Gaussian processes work. This design philosophy prioritizes transparency and usability over exposing optimization complexity. Future work could further simplify the experience through progressive disclosure techniques~\cite{springer_progressive_2018, eiband_bringing_2018}, where advanced concepts are revealed only when designers choose to explore them, making the system even more accessible to practitioners without optimization backgrounds.
	
Together, our findings demonstrate that multi-source optimization is not primarily about producing a single best outcome, but about structuring exploration in ways that surface heterogeneity, expose tensions, and support informed choices. Our findings highlight multi-source feedback as a productive lens for examining how designers engage with diverse evaluative signals. When designing such systems, the goal should be to integrate new feedback sources without disrupting established workflows. Extending existing human-in-the-loop optimization practices in this lightweight way enables designers, without requiring optimization expertise, to navigate alternatives and reconcile differing inputs. By showing that Bayesian optimization can accommodate this diversity with minimal adaptation, we contribute to a broader understanding of optimization not merely as a tool for convergence, but as a mechanism for reflection, negotiation, and collaboration across evaluative perspectives in design practice.

\section{Limitations and Future Work}
\label{sec:limitations}
Our findings also reveal open challenges in supporting multi-source feedback in design. Study~6 showed that both single- and multi-surrogate approaches have difficulty when optima shift over time, pointing to a general challenge in applying Bayesian optimization in non-stationary settings. In the empirical evaluation, designers encountered a related issue: although simulator feedback was useful, its slow response and opaque metrics limited how effectively it could inform design decisions.

These and other observations suggest several avenues for further study. Adaptive strategies could help surrogates accommodate changing objectives, and acquisition functions that incorporate source disagreement, such as a multi-source variant of Upper Confidence Bound, could improve exploration. While our approach gives designers control over source weighting, future work could explore automatically detecting when reweighting is beneficial and visually indicating these moments through progressive disclosure techniques. Enhancing source transparency and responsiveness, and conducting larger controlled studies with objective metrics like task efficiency or design quality, remain important next steps. Future work must also systematically validate whether and how designers dynamically manage, add, or reweight sources within real, unconstrained design workflows.

\section{Conclusion}
\label{sec:conclusion}
This paper examined how Bayesian optimization can support design scenarios where multiple heterogeneous sources contribute to evaluation. By adapting Bayesian optimization to model each source independently and combining their predictions into a unified decision process, we provided a simple and accessible way to integrate diverse signals. Embedding this approach into a no-code interface demonstrates a practical environment where designers can interact with multiple sources, laying the groundwork for workflows where feedback can be adjusted, added, or removed without disrupting the optimization process.

Through technical evaluations, we found that this independent modeling approach preserves optimization performance across a variety of complex, heterogeneous source settings. Furthermore, our formative user study provided insights into how designers navigate multi-source feedback when exploring visual alternatives within an interactive interface. Participants reported a perception of enhanced flexibility and confidence when working with the combined feedback. While our algorithmic evaluations demonstrate the technical capabilities of this system under dynamic conditions, future work is required to systematically evaluate how designers engage with these interactive mechanisms and how they impact final design outcomes in unconstrained workflows.

Overall, our work highlights multi-source Bayesian optimization as a viable approach for human-in-the-loop design tools. By providing both the independent modeling adaptation and a functional interface template, we demonstrate how computational systems can accommodate multiple evaluative signals without disrupting human intent. Ultimately, this reframing positions Bayesian optimization not merely as a mechanism for automated convergence, but as an interactive tool that can support designers in navigating complex, multi-source environments.

\begin{acks}
This work was supported by the Research Council of Finland project Subjective Functions (grant 357578), Finnish Center for Artificial Intelligence (grants 328400, 345604, 341763), European Research Council Advanced Grant (no. 101141916), the Department of Information and Communications Engineering at Aalto University.
\end{acks}

\bibliographystyle{ACM-Reference-Format}
\bibliography{references}

\appendix

\section{Technical Evaluation Details}
\label{app:technical-evaluation}

\paragraph{Ground-truth (GT) rationale.}
Our GT families were selected to reflect both challenges in design practice and well-established optimization testbeds.  

\emph{Sinusoidal mixtures} (Studies~1–2) represent settings where simulators or analytic proxies produce many evaluations while human or expert feedback is sparse or covers only part of the design space. Their smooth, interpretable structure makes GP fits and disagreements directly visible, supporting analysis of how multi-surrogate modeling handle mismatched availability and partial coverage.  

\emph{Classical 2D benchmarks} (Rosenbrock, Three-Hump Camel, Booth; Studies~3–6) are used as mathematical proxies for design project dynamics, such as shifting priorities and evolving definitions of quality. From an optimization perspective, these functions are standard stress-tests: Rosenbrock exposes ill-conditioned curved valleys, Three-Hump Camel adds multimodality, and Booth introduces coupled quadratic structure~\cite{sharma_metaheuristic_2024,Naser2025Survey}. With shifts and weightings, they provide controllable changes of optima, enabling systematic evaluation of adding/removing sources, mid-run reweighting, scaling to many sources, and non-stationary objectives.  

\paragraph{Evaluation Protocol.}
All studies rely on multiple randomized instances, 100 repetitions, and fixed modeling budgets. Known optima allow objective evaluation, ensuring results are robust, interpretable, and reproducible.

\subsection{Study 1: Multi-Pace Sampling}

This study evaluates how multi-surrogate modeling performs when sources differ in their sampling densities and observation noise. The objective is the average of sine and cosine,
\[
F_{\text{true}}(x) = \tfrac{1}{2}\big(\sin(x)+\cos(x)\big),
\quad x \in [0,10].
\]

\paragraph{Source definitions}
We define three sources $S_1$, $S_2$, and $S_3$ with distinct sampling densities and functional forms. All are sampled uniformly over $[0,10]$:
\begin{itemize}
	\item $S_1$: 100 samples, functional form $0.7\sin(x)+0.0\cos(x)$.
	\item $S_2$: 10 samples, functional form $0.4\sin(x)+0.6\cos(x)$.
	\item $S_3$: 20 samples, functional form $0.2\sin(x)+0.8\cos(x)$.
\end{itemize}

\paragraph{Gaussian process models}
Each surrogate is modeled with a GP using a Matérn kernel with $\nu=2.5$, length-scale initialized to $1.0$, and a constant kernel with $\theta=1.0$ (optimized in $[10^{-3},10^{3}]$). The GP noise variance is fixed to the squared observation noise of each source, determined by the regime under evaluation.

\paragraph{Single-surrogate baseline}
The baseline uses only the sparsest sampling rate (10 points). At these locations, all three sources are evaluated, and their values are averaged to form training data. A single GP is then fitted with effective noise variance equal to the average of the three source variances.

\paragraph{Multi-surrogate approach}
The multi-surrogate model maintains independent GPs for $S_1$, $S_2$, and $S_3$ at their native sampling rates. Predictions are combined by averaging the GP means and uncertainties.

\paragraph{Evaluation}
All models are evaluated on a grid of 200 points over $[0,10]$. Mean squared error (MSE) is computed against $F_{\text{true}}(x)$. To probe sensitivity to observation quality, we test six distinct noise regimes:
\begin{itemize}
	\item \textbf{Default:} $\sigma_1=0.20$, $\sigma_2=0.05$, $\sigma_3=0.10$.
	\item \textbf{Noisy $S_1$:} $\sigma_1=0.40$, $\sigma_2=0.05$, $\sigma_3=0.10$.
	\item \textbf{Noisy $S_2$:} $\sigma_1=0.20$, $\sigma_2=0.15$, $\sigma_3=0.10$.
	\item \textbf{Noisy $S_3$:} $\sigma_1=0.20$, $\sigma_2=0.05$, $\sigma_3=0.25$.
	\item \textbf{All low:} $\sigma_1=0.05$, $\sigma_2=0.02$, $\sigma_3=0.05$.
	\item \textbf{All high:} $\sigma_1=0.30$, $\sigma_2=0.20$, $\sigma_3=0.25$.
\end{itemize}

\paragraph{Repeated runs and robustness}
Each noise regime is repeated over $100$ random seeds. For every run we compute the MSE of the single- and multi-surrogate approaches, as well as their difference. The win rate, i.e., the percentage of seeds where the multi-surrogate outperforms the single-surrogate baseline, provides a compact measure of robustness. Results show that the multi-surrogate achieves its clearest advantage when $S_1$ is noisy (72\% win rate). Performance is more balanced in the \emph{All low}, \emph{Default}, and \emph{Noisy $S_3$} regimes (65\%), while the advantage decreases in the \emph{Noisy $S_2$} (54\%) and \emph{All high} (57\%) regimes.

\subsection{Study 2: (Dis)Agreement Between Sources}
In this study we evaluate disagreement between surrogates when sources observe only part of the parameter domain. This setting reflects cases where some evaluation sources cannot assess all designs.  

\paragraph{Synthetic functions}  
We define a ground truth function
\[
f(x) = \frac{1}{2}(\sin(x) + \cos(x)),
\]
and three sources with different linear combinations of $\sin(x)$ and $\cos(x)$:
\[
S_1(x) = \sin(x), \quad
S_2(x) = 0.2\sin(x) + 0.8\cos(x), \quad
S_3(x) = 0.4\sin(x) + 0.6\cos(x).
\]

\paragraph{Sampling setup}  
The evaluation domain is $[0,6]$.  
\begin{itemize}
	\item Source~1 covers the entire domain with 15 samples, corrupted with Gaussian noise $\mathcal{N}(0, 0.15^2)$.  
	\item Source~2 covers only the subdomain $[0,3]$ with 8 samples, corrupted with Gaussian noise $\mathcal{N}(0, 0.10^2)$.  
	\item Source~3 covers only the subdomain $[3,6]$ with 8 samples, corrupted with Gaussian noise $\mathcal{N}(0, 0.05^2)$.  
\end{itemize}

\paragraph{Model fitting}  
For each source we fit a GP with kernel
\[
k(x,x') = \theta \, \mathrm{Matern}_{\nu=2.5}(\lVert x - x' \rVert; \ell=1.0),
\]
where the constant scaling parameter $\theta$ is initialized to $1.0$ and optimized during training within the range $[10^{-3}, 10^{3}]$. The length-scale $\ell$ is initialized to $1.0$ and optimized without fixed bounds. Targets are normalized. The noise variance parameter $\alpha$ is fixed to the square of the sampling noise for each source. GPs are trained on the sampled data and evaluated on a dense grid of 200 test points in $[0,6]$.  

\paragraph{Combined surrogate}  
To obtain a simple aggregate, we compute the mean of the three GP predictions at each test point:
\[
\mu_{\text{comb}}(x) = \tfrac{1}{3} \sum_{i=1}^3 \mu_i(x), \quad
\sigma^2_{\text{comb}}(x) = \tfrac{1}{3} \sum_{i=1}^3 \sigma_i^2(x).
\]

\paragraph{Quantifying disagreement}  
Disagreement is measured as the variance of the predicted means across surrogates:
\[
D(x) = \mathrm{Var}\!\left(\mu_1(x), \mu_2(x), \mu_3(x)\right).
\]
To separate disagreement from model uncertainty, we also compute the average predictive variance across surrogates,
\[
\bar{\sigma}^2(x) = \tfrac{1}{3} \sum_{i=1}^3 \sigma_i^2(x).
\]
This distinction highlights regions where sources genuinely disagree versus regions where predictions are uncertain due to limited data.

\subsection{Study 3: Runtime Adjustment of Sources}

This study evaluates how multi-surrogate modeling adapts when evaluation sources are added or removed during the optimization. We simulate a four-phase scenario with two-dimensional test functions.

\paragraph{Source definitions}
We define three sources, each given by a shifted analytic benchmark function with Gaussian observation noise:
\begin{itemize}
	\item $S_1$: Rosenbrock function,
	\[
	f_{\text{rosenbrock}}(x_0,x_1) = (1-x_0)^2 + 100\,(x_1 - x_0^2)^2,
	\]
	with noise $\sigma_1 = 0.15$.
	\item $S_2$: Three-Hump Camel function,
	\[
	f_{\text{camel}}(x_0,x_1) = 2x_0^2 - 1.05x_0^4 + \tfrac{1}{6}x_0^6 + x_0x_1 + x_1^2,
	\]
	with noise $\sigma_2 = 0.10$.
	\item $S_3$: Booth function,
	\[
	f_{\text{booth}}(x_0,x_1) = (x_0 + 2x_1 - 7)^2 + (2x_0 + x_1 - 5)^2,
	\]
	with noise $\sigma_3 = 0.05$.
\end{itemize}
Each source is randomly translated within $[-2,2]^2$ so that the location of its minimum differs across runs, ensuring independence between sources.

\paragraph{Phase schedule}
Optimization is divided into four phases of 12 steps each (48 steps total):
\begin{itemize}
	\item \textit{Phase 1:} $S_1$ only.
	\item \textit{Phase 2:} $S_1 + S_2$.
	\item \textit{Phase 3:} $S_1 + S_2 + S_3$.
	\item \textit{Phase 4:} $S_1 + S_3$ (removing $S_2$).
\end{itemize}

At each phase, only the active sources are evaluated. 

\paragraph{Gaussian process models}
Each source is modeled by an independent GP with kernel
\[
k(x,x') = \theta \cdot \mathrm{Matern}_{\nu=2.5}(\lVert x-x'\rVert),
\]
where $\theta$ is initialized to $1.0$ and optimized in $[10^{-3},10^3]$. The GP noise variance is fixed to $\sigma_i^2$ for source $S_i$.  

The multi-surrogate model aggregates active surrogates by averaging their predictive means. Baselines use a single GP trained on fused labels.

\paragraph{Optimization and comparison}
We run 100 experiments with Optuna and the TPE sampler~\cite{akiba2019optuna} allocating 12 trials per phase (50 total trials). Three methods are compared:
\begin{itemize}
	\item \textbf{Multi-Surrogate (ours):} independent GPs per source, retained across phases.
	\item \textbf{Single -- Retains data:} restart a single GP at each phase using all past evaluations under the active sources.
	\item \textbf{Single -- Discards data:} restart a single GP at each phase, discarding all past data.
\end{itemize}

\paragraph{Evaluation}
Performance is measured by two metrics at each phase:
\begin{itemize}
	\item The average true objective value of the sampled candidates.
	\item The best (lowest) true objective value so far.
\end{itemize}
The global minimum of the average of the three underlying functions is computed by grid search over $[-3,3]^2$ and shown as a reference.

\subsection{Study 4: Mid-Optimization Reweighting}
\label{app-sec:reweighting}
This study measures how the optimizer adapts when the relative weights of two sources change during the run. We consider three cases where the fused optimum moves by a small, medium, or large distance.

\paragraph{Problem setting}
We optimize a 1D design variable $x \in [0,10]$. The fused objective is
\[
y(x;t) \;=\; w_1(t)\,S_1(x) \;+\; w_2(t)\,S_2(x),
\]
with a single reweighting at step $t{=}21$. Total budget is $T{=}40$ steps per run. We run $N_{\text{runs}}{=}100$ independent repetitions. Optimization uses Optuna with the TPE sampler and $n_{\text{startup}}{=}5$ startup trials. For reporting and internal calculations we use a uniform grid of 2000 points on $[0,10]$.

\paragraph{Source means and noises}
Each case defines smooth functions $S_1$ and $S_2$ (quadratic baseline plus a negative Gaussian) and Gaussian observation noise:
\begin{itemize}
	\item \textbf{Small-distance case:}  
	\[
	\begin{aligned}
		S_1^\text{mean}(x) &= 0.03\,(x-5.0)^2 \;-\; 1.10\,\exp\!\Big(-\tfrac{1}{2}\big(\tfrac{x-5.0}{0.50}\big)^2\Big),\\
		S_2^\text{mean}(x) &= 0.03\,(x-5.0)^2 \;-\; 1.00\,\exp\!\Big(-\tfrac{1}{2}\big(\tfrac{x-5.6}{0.55}\big)^2\Big),
	\end{aligned}
	\]
	noise levels $(\sigma_1,\sigma_2)=(0.08,\,0.05)$, weights $w_{\text{pre}}=(0.55,\,0.45)$ and $w_{\text{post}}=(0.45,\,0.55)$.
	\item \textbf{Medium-distance case (main paper):}  
	\[
	\begin{aligned}
		S_1^\text{mean}(x) &= 0.05\,(x-5.0)^2 \;-\; 1.30\,\exp\!\Big(-\tfrac{1}{2}\big(\tfrac{x-3.2}{0.40}\big)^2\Big),\\
		S_2^\text{mean}(x) &= 0.05\,(x-5.0)^2 \;-\; 1.35\,\exp\!\Big(-\tfrac{1}{2}\big(\tfrac{x-7.1}{0.45}\big)^2\Big),
	\end{aligned}
	\]
	$(\sigma_1,\sigma_2)=(0.10,\,0.05)$, $w_{\text{pre}}=(0.60,\,0.40)$, $w_{\text{post}}=(0.35,\,0.65)$.
	\item \textbf{Large-distance case:}  
	\[
	\begin{aligned}
		S_1^\text{mean}(x) &= 0.06\,(x-5.0)^2 \;-\; 1.80\,\exp\!\Big(-\tfrac{1}{2}\big(\tfrac{x-1.5}{0.28}\big)^2\Big),\\
		S_2^\text{mean}(x) &= 0.06\,(x-5.0)^2 \;-\; 1.85\,\exp\!\Big(-\tfrac{1}{2}\big(\tfrac{x-8.6}{0.30}\big)^2\Big),
	\end{aligned}
	\]
	$(\sigma_1,\sigma_2)=(0.12,\,0.05)$, $w_{\text{pre}}=(0.80,\,0.20)$, $w_{\text{post}}=(0.20,\,0.80)$.
\end{itemize}
Noisy observations are $S_i(x) = S_i^\text{mean}(x) + \varepsilon_i$ with $\varepsilon_i \sim \mathcal{N}(0,\sigma_i^2)$.

\paragraph{Methods}
\begin{enumerate}[label=(\Alph*),itemsep=2pt,topsep=2pt]
	\item \textbf{Single -- No history}: restart the optimization at $t{=}21$ and discard all past data, i.e., the new optimization begins from scratch under $w_{\text{post}}$.
	\item \textbf{Single -- Past history}: restart the optimization at $t{=}21$, but reuse the past fused labels computed under $w_{\text{pre}}$ as initialization for the new optimization under $w_{\text{post}}$.
	\item \textbf{Single -- Reweighted history}: restart the optimization at $t{=}21$, and reuse the past evaluations with their labels recomputed under $w_{\text{post}}$ as initialization to the optimizer.
	\item \textbf{Multi -- Reweighted history plus EI warm-start (K=20)}: fit two independent GPs on pre-change $(x, S_1)$ and $(x, S_2)$ data; form a fused posterior  
	$\mu_f(x)=w_1\mu_1(x)+w_2\mu_2(x)$ and  
	$\sigma_f^2(x)=w_1^2\sigma_1^2(x)+w_2^2\sigma_2^2(x)$.  
	Compute expected improvement (EI) for minimization,
	\[
	\mathrm{EI}(x) \;=\; (f_{\text{best}}-\mu_f(x))\,\Phi(z) \;+\; \sigma_f(x)\,\phi(z),\quad
	z=\frac{f_{\text{best}}-\mu_f(x)}{\sigma_f(x)}.
	\]
	Select Top-$K$ EI peaks with a suppression radius of $2\%$ of the domain. Use these points, together with the reweighted past data, as initialization to the optimizer. The main comparison uses $K{=}20$; we also sweep $K \in \{20,50,100,200,300\}$.
\end{enumerate}

\begin{figure}[t]
	\centering
	\includegraphics[width=0.9\textwidth]{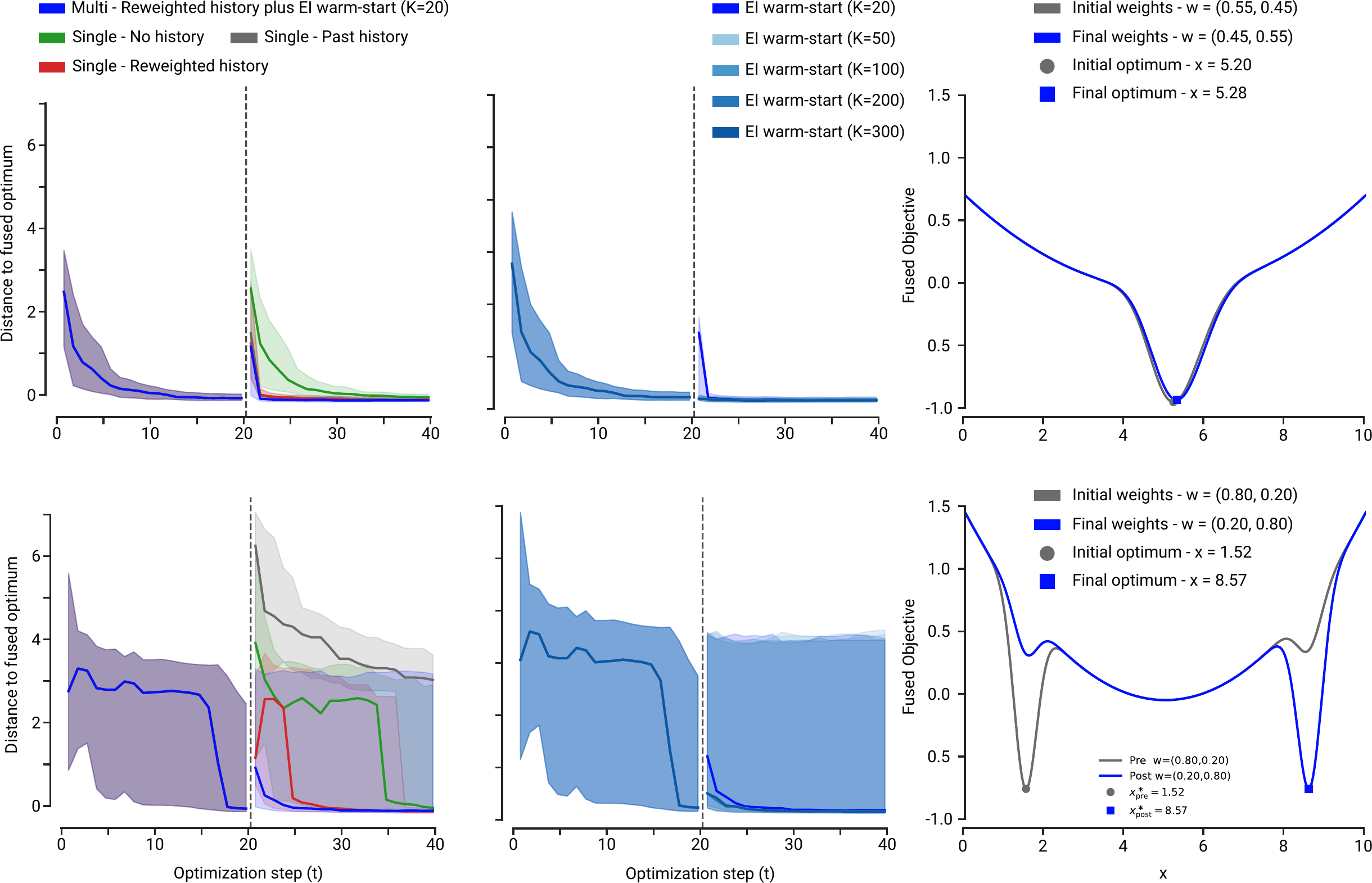}
	\caption{Mid-optimization reweighting (Study~4 Appendix). The dashed line marks the point at $t{=}21$ when source weights change and the optimizer restarts.  
		\textbf{Left:} comparison of the three single-surrogate baselines with the multi-surrogate strategy (\emph{Multi -- Reweighted history plus EI warm-start}, $K{=}20$).  
		\textbf{Center:} effect of varying the number of EI initialization points in the multi-surrogate method.  
		\textbf{Right:} fused source minima before and after reweighting.  
		Results are shown for the small- (top row) and large-distance (bottom row) cases. The medium-distance case appears in the main paper.}
	\Description{Six plots showing results of Study 4 on mid-optimization reweighting. 
		The left column compares optimization performance across four methods. 
		The multi-surrogate method (blue: Reweighted history plus EI warm-start, K=20) is contrasted with three single-surrogate baselines: 
		green (Single – No history), gray (Single – Past history), and red (Single – Reweighted history). 
		The y-axis shows distance to the fused optimum and the x-axis shows optimization steps. 
		In both the small-distance case (top row) and large-distance case (bottom row), the multi-surrogate method converges faster and more reliably, especially after reweighting at step 21 (marked by a dashed vertical line). 
		The center column shows sensitivity of the multi-surrogate method to the number of EI warm-start points (20, 50, 100, 200, 300). Convergence is best with fewer initialization points, and performance degrades as the number increases. 
		The right column plots fused objective curves before and after reweighting. In the small-distance case, the minima shift slightly, while in the large-distance case the shift is substantial, moving the optimum across the domain. 
		Together the plots illustrate that the proposed multi-surrogate method adapts more robustly to mid-run changes in source weights than any of the single-surrogate baselines.}
	\label{fig:study4-small-large}
\end{figure}

The supplementary results for the small- and large-distance cases are reported in Fig.~\ref{fig:study4-small-large}.

\paragraph{Gaussian process details}
Each source GP is a scikit-learn \texttt{GaussianProcessRegressor} with kernel
\[
k(x,x') \;=\; \theta \cdot \mathrm{Matern}_{\nu=2.5}(\lvert x-x'\rvert),
\]
where $\theta$ is initialized to $1.0$ and optimized in $[10^{-3},10^{3}]$. Targets are normalized. The optimizer is L-BFGS-B. The GP noise parameter is fixed to the known observation variance.

\paragraph{Metric}
We report the Euclidean distance to the current fused optimum
\[
d_t \;=\; \big\lvert x_{\text{best},t} - x_t^\ast \big\rvert,
\]
where $x_t^\ast$ is the minimizer of the \emph{noise-free} synthetic fused mean under the current weights at step $t$. The best-so-far is reset at $t{=}21$.

\subsection{Study 5: Increasing the Number of Sources}

This study evaluates the effect of increasing the number of evaluation sources on optimization performance.

\paragraph{True objective}
The ground-truth function is defined as the mean of $M{=}15$ shifted Rosenbrock functions:
\[
F_{\text{true}}(x) = \frac{1}{M} \sum_{j=1}^{M} G_j(x),
\quad
G_j(x) = (1 - (x_1 - \delta_{j1}))^2 + 100 \,\big((x_2 - \delta_{j2}) - (x_1 - \delta_{j1})^2\big)^2,
\]
where $(\delta_{j1},\delta_{j2}) \in [-2,2]^2$ are random shifts assigned once per experiment.

\paragraph{Source definitions}
Each source $S_i$ is a weighted combination of the same $M$ shifted Rosenbrock functions:
\[
S_i(x) = \sum_{j=1}^{M} w_{ij}\, G_j(x),
\]
with weights $w_{ij}$ drawn uniformly in $[0,1]$ and resampled for every run. This makes sources distinct from one another and from $F_{\text{true}}$.

\paragraph{Experimental design}
We vary the number of sources as $N \in \{1,2,5,8,10\}$. Two modeling strategies are compared:
\begin{itemize}
	\item \textbf{Single-surrogate baseline}: aggregate all sources into one GP by averaging their outputs.
	\item \textbf{Multi-surrogate}: fit one GP per source and fuse predictions by averaging means.
\end{itemize}
All GPs use the kernel
\[
k(x,x') = \theta \cdot \mathrm{Matern}_{\nu=2.5}(\lVert x-x'\rVert),
\]
with $\theta$ initialized to $1.0$ and optimized in $[10^{-3},10^{3}]$, and noise variance fixed to $10^{-6}$. Optimization is performed with Optuna and the TPE sampler, with $10$ random initial points and $50$ optimization steps.

\paragraph{Evaluation}
Each condition is repeated $100$ times. Performance is measured by the Euclidean distance between the optimizer's best candidate $\hat{x}$ and the precomputed minimizer $x^{\ast}$ of $F_{\text{true}}$, obtained via grid search over $[-2,2]^2$ with resolution $300 \times 300$.

\subsection{Study 6: Non-Stationary Optimization}

This study evaluates how single- and multi-surrogate Bayesian optimization methods perform when the underlying objective changes during optimization.

\paragraph{True objective}
We define a non-stationary ground truth function based on the Rosenbrock benchmark:
\[
F_{\text{true}}(x,t) = (1 - (x_1 - s_1(t)))^2 + 100 \,\big((x_2 - s_2(t)) - (x_1 - s_1(t))^2\big)^2,
\]
where $x = (x_1, x_2)$ and $s(t) \in \mathbb{R}^2$ is a time-dependent shift schedule. The shift $s(t)$ changes every $10$ optimization steps, producing $5$ stationary blocks of $10$ steps each. At trial $t$, the global minimum is located at $s(t) + (1,1)$.

\paragraph{Source definitions}
We use $N=3$ sources, each defined as a biased variant of the non-stationary Rosenbrock:
\[
S_i(x,t) = (1 - (x_1 - s_1(t) - b_{i1}))^2 + 100 \,\big((x_2 - s_2(t) - b_{i2}) - (x_1 - s_1(t) - b_{i1})^2\big)^2,
\]
where $b_i = (b_{i1}, b_{i2})$ is a fixed random offset sampled uniformly from $[-1,1]^2$ at the start of each run and kept constant throughout.

\paragraph{Experimental design}
Optimization runs consist of $n_\text{steps}=50$ trials, divided into 5 blocks of 10 trials each. The shift schedule is initialized at $s(0) = (-1,-1)$ and updated by $(+0.5,+0.5)$ at the start of every new block. Both single- and multi-surrogate methods are initialized with $n_\text{init}=10$ random evaluations and proceed for $n_\text{trials}=50$ optimization steps using Optuna with the TPE sampler. Each condition is repeated $100$ times with shared source biases across methods to ensure comparability.  

\paragraph{Modeling strategies}
\begin{itemize}
	\item \textbf{Single-surrogate}: a single GP is trained on the average of the three source values at each $x$.  
	\item \textbf{Multi-surrogate}: three independent GPs are maintained, one per source, and predictions are fused by averaging their means.  
\end{itemize}
Both strategies use Matérn kernels with $\nu=2.5$, length scale initialized to $1.0$ and optimized in $[10^{-3},10^{3}]$, and noise variance fixed to $10^{-6}$.

\paragraph{Evaluation}
Performance is measured per block using two metrics:
\begin{enumerate}
	\item The \emph{mean distance}, i.e., the average Euclidean distance between all evaluated points in the block and the current optimum $x^\ast_t$.
	\item The \emph{best distance}, i.e., the minimum Euclidean distance achieved in the block.  
\end{enumerate}

\section{\toolname Walkthrough}
\label{app:walkthrough}
\begin{figure}[t]
	\centering
	\begin{subfigure}{\linewidth}
		\centering
		\includegraphics[width=\linewidth]{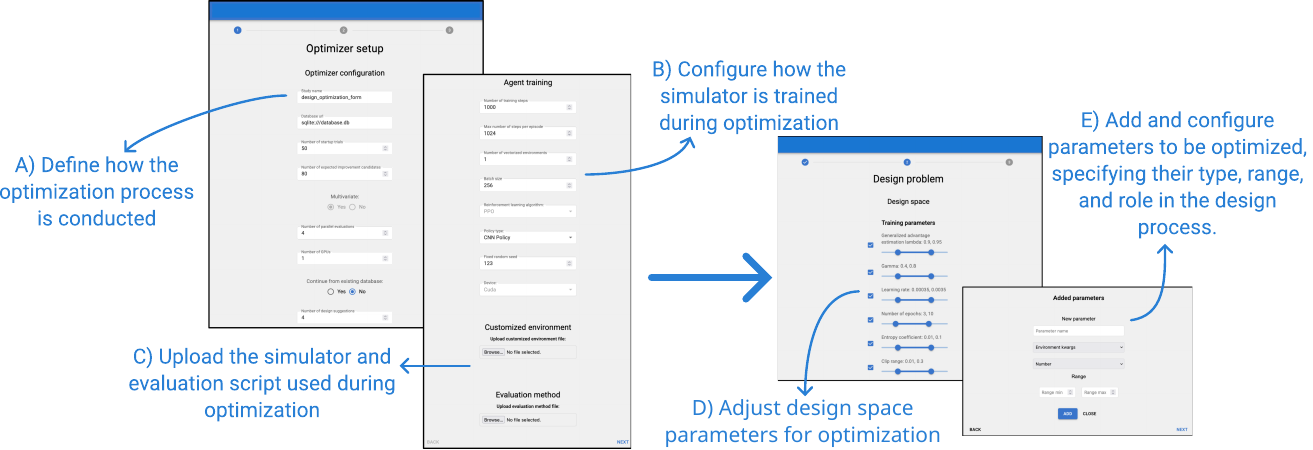}
		\caption{Interface for configuring the optimization process in \toolname, including optimizer setup, simulator training, evaluation scripts, and design space parameters.}
		\Description{Screenshot of the Muse interface for configuring optimization. 
			The interface is divided into labeled areas. 
			A) lets users define the optimization process, including optimizer setup and budgets. 
			B) configures how the simulator is trained, with learning settings and environment options. 
			C) is used to upload the simulator and evaluation scripts. 
			D) allows adjustment of design space parameters such as ranges and variable types. 
			E) supports adding and configuring parameters to be optimized, specifying their role in the design process. 
			Arrows indicate the sequential flow from defining the optimization setup, through simulator training, to evaluation and design space configuration.}
		\label{fig:ui_walkthrough-app}
	\end{subfigure}
	
	\vspace{1em}
	
	\begin{subfigure}{0.7\linewidth}
		\centering
		\includegraphics[width=\linewidth]{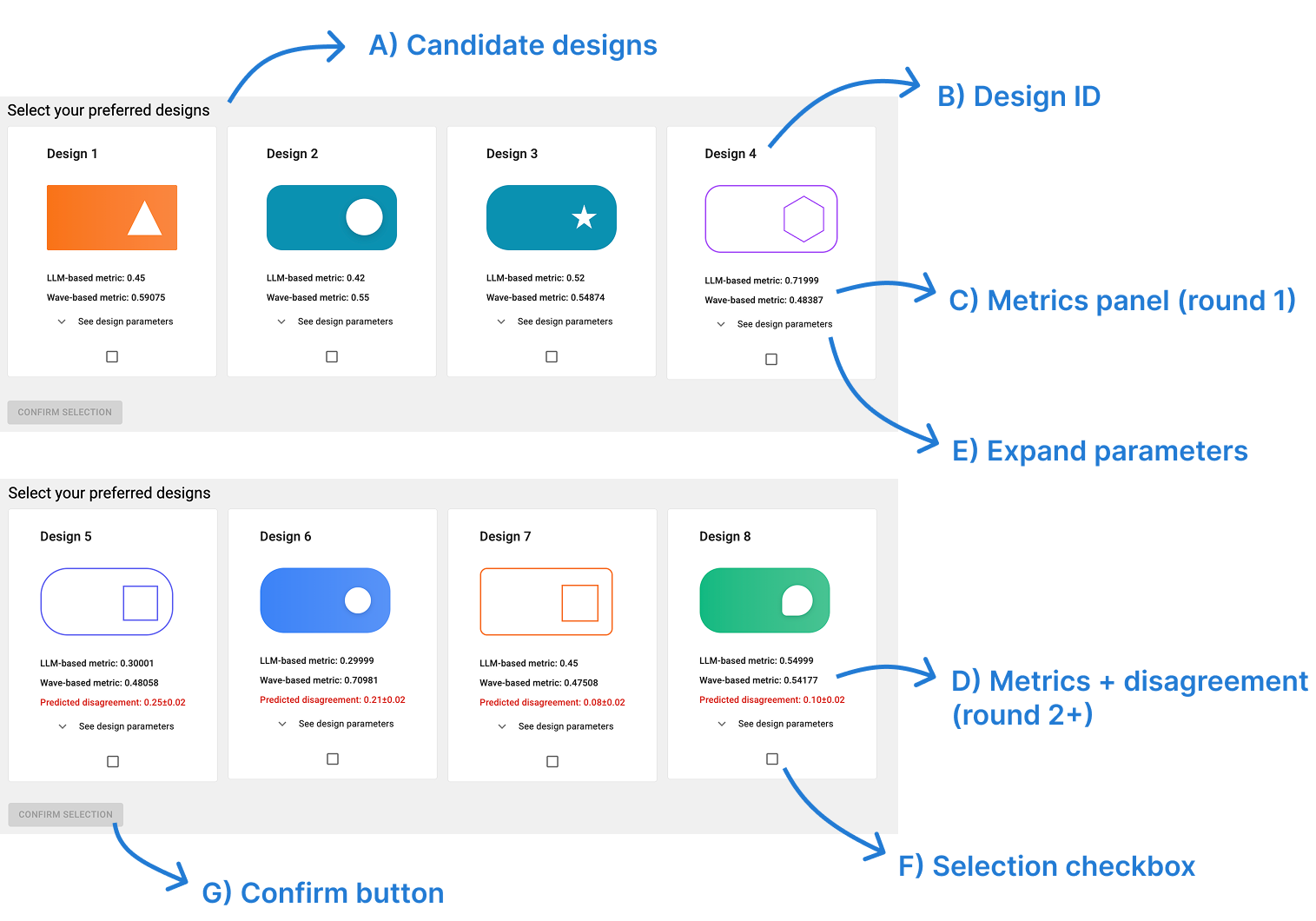}
		\caption{Interface for selecting preferred designs in \toolname, showing candidate designs, evaluation metrics, disagreement predictions, and selection controls.}
		\Description{Screenshot of the Muse interface for selecting preferred designs. 
			Candidate designs are shown as individual cards with labels A to G. 
			A) Candidate designs display visual representations of each option. 
			B) Each card shows a design identifier. 
			C) In the first round, numerical metrics describing the design are displayed. 
			D) In later rounds, metrics also include disagreement values across sources. 
			E) An expand option reveals additional parameter details. 
			F) A checkbox lets the user select preferred designs. 
			G) A confirm button appears below the cards to finalize the selection.}
		\label{fig:task_flow}
	\end{subfigure}
	
	\caption{User interface of \toolname. 
		(a) Configuration of the optimization process, including setup, simulator training, evaluation scripts, and design space parameters. 
		(b) Selection of preferred designs, with candidate visualization, evaluation metrics, disagreement predictions, and selection controls.}
	\label{fig:ui_combined}
\end{figure}

Figures~\ref{fig:ui_walkthrough-app} and~\ref{fig:task_flow} illustrate the user interface of \toolname. 
The system conceptually supports the integration of simulators implemented as reinforcement learning-based agents, which can be configured and trained directly within the optimization workflow. 
The first figure shows how users set up the optimizer, upload simulator and evaluation scripts, and define the design space. 
The second figure demonstrates how candidate designs are presented for user selection, along with evaluation metrics and disagreement estimates across sources.

\section{Details of the Toggle Demonstration Task}

\subsection{Design Task}
\begin{figure}[t]
	\centering
	\includegraphics[width=0.9\linewidth]{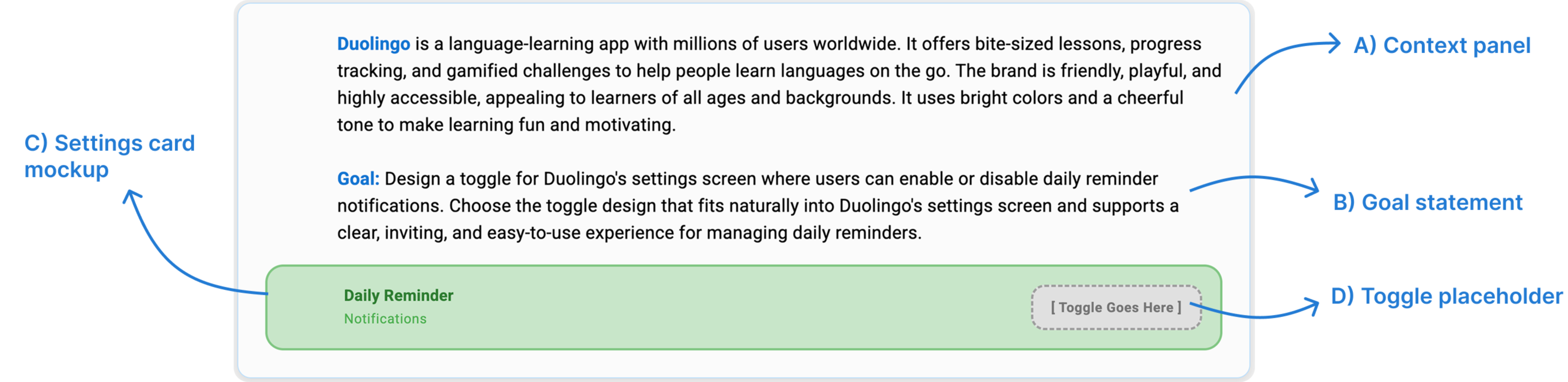}
	\caption{Illustration of a design task prompt for creating a toggle for Duolingo’s notifications. A) Short description of the product and brand cues (playful, inviting, accessible) to set the intended tone. B) The concrete task: "Design a settings-screen toggle for daily reminders." C) The visual surface where the toggle will live (title “Daily Reminder” with a supporting subtitle). D) Target placement indicating where the designed toggle appears ("[Toggle Goes Here]").}
	\Description{Illustration of a design task prompt for creating a toggle in Duolingo’s settings screen. 
		At the top, a context area describes the Duolingo app as playful, inviting, and accessible, highlighting its bright colors and motivating tone. 
		Below, a goal statement instructs participants to design a toggle that enables or disables daily reminder notifications, fitting naturally into the settings screen and supporting ease of use. 
		A settings card mockup is shown with the text “Daily Reminder – Notifications,” and a highlighted placeholder labeled “[Toggle Goes Here]” marks the target location where the designed toggle should be placed. 
		Arrows label the four elements: A) context description, B) goal statement, C) settings card mockup, and D) toggle placeholder.}
	\label{fig:design-brief}
\end{figure}

The design task, shown in \Fig{fig:design-brief}, was a Duolingo-inspired one where the student designed the settings toggle for enabling daily reminder notifications. The task emphasizes a tone that is cheerful, approachable, and highly accessible, consistent with Duolingo's  brand. The toggle should naturally fit within a settings card layout. The interaction was hover-triggered, allowing participants to preview the transition while choosing among toggles.

\subsection{Design Parameters}
The toggle design is based on the seven parameters: animation duration, border radius, thumb size ratio, thumb shape, easing curve, visual style and color scheme. Hover is the interaction method, which triggers state changes and applies to all generated toggles. Animation duration controls how fast the toggle shifts between the on and off states, while border radius adjusts the curvature of the track. Thumb size ratio defines the size of the thumb relative to the track. Next is the thumb shape which includes options circle, square, diamond, hexagon, star, triangle, teardrop and bean.  The easing curve has options of standard CSS easing functions linear, ease, ease-in, ease-out, ease-in-out. The visual style affects both track and thumb. The options for visual style are flat-solid, shadow-gradient, glow-glass, outline and elevated-textured each having a corresponding style. And the last parameter color scheme defines the palette colors used for the active track, inactive track and thumb. The choices are modern blue, nature green, sunset orange, dark mode, neon purple and enterprise gray. Together, those parameters and their different combinations give a considerable design space to explore desired toggle designs. For optimization, categorical parameters were internally mapped to zero-indexed integers, which were then decoded back into their corresponding visual styles during rendering. See Table \ref{tab:toggle-params} for parameter ranges and option sets.

\begin{table}[t]
	\centering
	\caption{Toggle design parameters with ranges and option sets.}
	\Description{Table of toggle design parameters, their ranges or option sets, and short descriptions. 
		Parameters include: 
		Animation Duration, ranging from 100 to 1000 milliseconds, describing the speed of transitions; 
		Border Radius, 0 to 50 pixels, controlling track curvature; 
		Thumb Size Ratio, 0.30 to 0.95, defining thumb size relative to track; 
		Thumb Shape, with options Circle, Square, Diamond, Hexagon, Star, Triangle, Teardrop, and Bean, describing the geometry of the thumb; 
		Easing Curve, with options Linear, Ease, Ease-In, Ease-Out, and Ease-In-Out, describing animation motion profile; 
		Visual Style, with options Flat Solid, Shadow Gradient, Glow Glass, Outline Style, and Elevated Textured, defining the overall appearance; 
		and Color Scheme, with options Modern Blue, Nature Green, Sunset Orange, Ocean Teal, Dark Mode, Neon Purple, and Enterprise, defining predefined palettes for active track, inactive track, and thumb.}
	\label{tab:toggle-params}
	\begin{tabular}{p{3cm} p{5cm} p{4.5cm}}
		\toprule
		\textbf{Parameter} & \textbf{Range / Options} & \textbf{Description} \\
		\midrule
		Animation Duration & 100–1000 ms & Speed of transition between on/off states \\
		Border Radius & 0–50 px & Curvature of the track edges \\
		Thumb Size Ratio & 0.30–0.95 & Relative size of the thumb compared to the track \\
		Thumb Shape & Circle, Square, Diamond, Hexagon, Star, Triangle, Teardrop, Bean & Geometry of the thumb element \\
		Easing Curve & Linear, Ease, Ease-In, Ease-Out, Ease-In-Out & Motion profile of the animation \\
		Visual Style & Flat Solid, Shadow Gradient, Glow Glass, Outline Style, Elevated Textured & Overall appearance of track and thumb \\
		Color Scheme & Modern Blue, Nature Green, Sunset Orange, Ocean Teal, Dark Mode, Neon Purple, Enterprise & Predefined palettes for active track, inactive track, and thumb \\
		\bottomrule
	\end{tabular}
\end{table}

\subsection{Rendering Details}
The rendering was done on a frontend using ReactJS via a fixed canvas of 200 by 100 pixels representing a React component called \texttt{AnimatedToggle} that mapped encoded parameter values to CSS styles. Polygonal clip-paths implemented shapes such as diamonds, hexagons, stars and triangles; other shapes used border-radius definitions. The changes of state occurred via CSS transforms with transitions based on the previously selected duration and easing.

\subsection{t-SNE and UMAP Visualizations of Toggle Designs}
\begin{figure}[!t]
	\centering
	\includegraphics[width=0.99\linewidth]{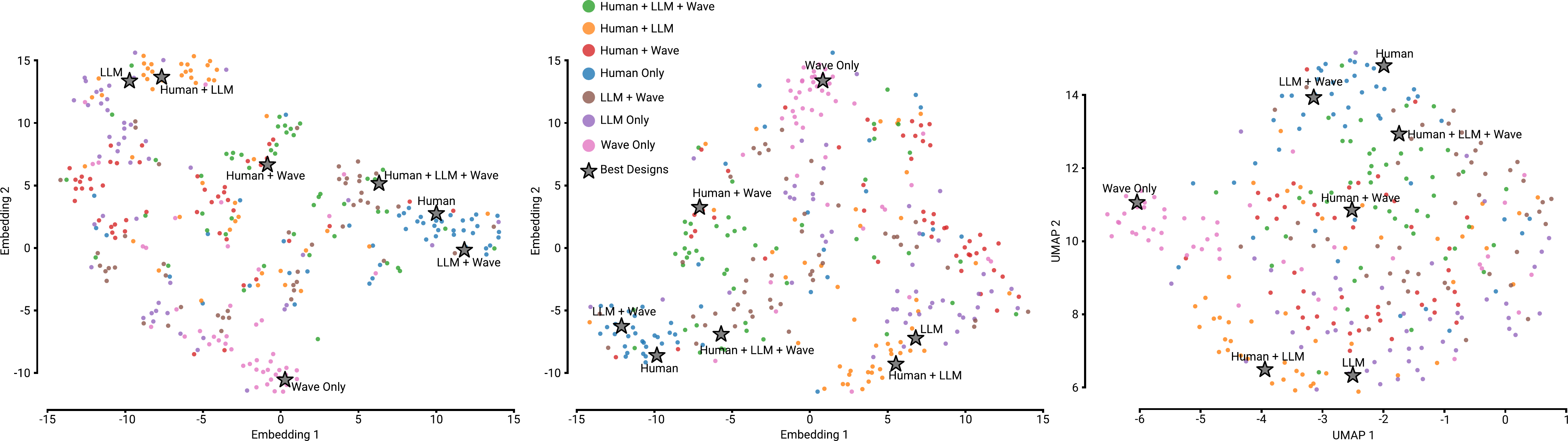}
	\caption{The toggle design demonstration shows how different source combinations shape both the search process and the resulting designs. 
		\textbf{Left} and \textbf{center}: t-SNE projections of toggle parameters with perplexity values of 10 and 50, respectively. Each dot represents a candidate evaluated during optimization, and distinct source combinations guide exploration into different regions of the design space. Dark stars mark the best designs identified under each condition. 
		\textbf{Right}: a UMAP projection of the same parameters, offering an alternative view of the design landscape.
	}
	\Description{Three scatterplots comparing how different evaluation source combinations influence toggle design optimization. 
		The left and center plots are t-SNE projections with different perplexity values (10 and 50, respectively), and the right plot is a UMAP projection. 
		Each dot represents a candidate toggle design, with colors indicating source combinations: green for Human+LLM+Wave, orange for Human+LLM, red for Human+Wave, blue for Human only, brown for LLM+Wave, purple for LLM only, and pink for Wave only. 
		Dark stars highlight the best designs identified under each condition, and are labeled with the corresponding source combination. 
		In all views, distinct clusters emerge where different source combinations guide exploration into different regions of the design space. 
		The UMAP projection provides an alternative layout of the same candidates, with similar clustering and clear separation between source-specific regions.
	}
	\label{fig:toggle-sm}
\end{figure}

\Fig{fig:toggle-sm} provides an extended visualization of the toggle design task. 
Two t-SNE projections (perplexity 10 and 50) and one UMAP projection show how different source combinations (Human, LLM, WAVE, and their combinations) guide the search into distinct regions of the design space. 
Dark stars mark the best solutions under each condition. 
These examples illustrate how feedback from different sources produces distinct visual and interactive outcomes, highlighting systematic differences in style, color, shape, and animation.

Along with the visualization task we introduced an interactive Toggle Task inspired by the toggle components commonly used in modern applications. This task served as a complementary design space where participants can experiment with and fine-tune UI components with various parameters. 

The toggle task served multiple objectives. It helped us refine users' preferences at a granular level and observe how well optimization works with complex, multi-parameter design spaces. It also allowed us to analyze how users evaluate elements in a setting that was visually intuitive but still customizable.

\subsection{LLM Evaluator Prompt for Toggle Demonstration}

For the toggle demonstration, we implemented an LLM evaluator (GPT-4o, Azure) instructed to act as a senior UI designer assessing toggle designs for Duolingo’s Daily Reminder feature. The model returned a score in \([0,1]\), where higher values reflected stronger brand alignment, clarity, and usability. The following shows the system and user prompts:

\begin{tcolorbox}[
	title=LLM Prompts,
	colback=gray!10,
	colframe=gray!40,
	boxrule=0.5pt,
	arc=2pt,
	outer arc=2pt,
	fonttitle=\bfseries,
	breakable
	]
	
	\textbf{System Prompt}
	
	You are a senior UI designer evaluating a toggle switch for Duolingo.  
	
	\textbf{Context — Design Task:}
	
	Duolingo is a language-learning app with millions of users worldwide.  
	It offers bite-sized lessons, progress tracking, and gamified challenges to help people learn languages on the go.  
	The brand is friendly, playful, and highly accessible, appealing to learners of all ages and backgrounds.  
	It uses bright colors and a cheerful tone to make learning fun and motivating.  
	
	\textbf{Goal:} Design a toggle for Duolingo's settings screen where users can enable or disable daily reminder notifications.  
	Choose the toggle design that fits naturally into Duolingo's settings screen and supports a clear, inviting, and easy-to-use experience for managing daily reminders.  
	
	\textbf{Output format:} return ONLY a JSON object: \verb|{"score": <number 0..1>}|  
	
	\textbf{How to interpret parameters:}  
	
	\textbullet{} Some fields are integer indices that must be decoded before judging.
	
	\quad\quad \textbullet{} \textbf{colorScheme} $\in$ \{0..6\} $\to$ modern, nature, sunset, ocean, dark, neon, enterprise  
	
	\quad\quad \textbullet{} \textbf{visualStyle} $\in$ \{0..4\} $\to$ flat, shadow, glow, outline, elevated  
	
	\quad\quad \textbullet{} \textbf{easing} $\in$ \{0..4\} $\to$ linear, ease, ease-in, ease-out, ease-in-out  
	
	\quad\quad \textbullet{} \textbf{thumbShape} $\in$ \{0..7\} $\to$ circle, square, diamond, hexagon, star, triangle, teardrop, bean  
	
	\textbullet{} \textbf{Non-index fields:  }
	
	\quad\quad \textbullet{} \textbf{duration} (ms): 200–400 ideal; $>$600 sluggish; $<$150 abrupt  
	
	\quad\quad \textbullet{} \textbf{borderRadius} (px): 10–35 modern; $>$45 bloated  
	
	\quad\quad \textbullet{} \textbf{thumbRatio} (0..1): 0.70–0.80 good balance  
	
	\quad\quad \textbullet{} \textbf{toggleSize} (px): 140–200 typical in this task  
	
	\textbf{Scoring guidelines (brand + UX):}  
	
	\quad\quad \textbullet{} \textbf{Brand fit:} playful, friendly, high-contrast  
	
	\quad\quad \textbullet{} \textbf{Clarity:} on/off state obvious, strong contrast  
	
	\quad\quad \textbullet{} \textbf{Motion:} easing consistent with mobile UI norms  
	
	\quad\quad \textbullet{} \textbf{Effects:} minimal glow, avoid gimmicks, outline acceptable with contrast  
	
	Return ONLY JSON with the numeric score.  
	
	Important: Scores for different designs must differ unless parameters are identical.  
	
	\textbf{User Prompt (example)}  
	
	Evaluate this toggle design for Duolingo. Decode any index parameters using the tables above, then score the design.  
	Return ONLY JSON with the field \verb|"score"|.  
	
	\textbf{Raw parameters:}
	\begin{verbatim}
		{
			"duration": 300,
			"borderRadius": 24,
			"thumbRatio": 0.75,
			"thumbShape": 0,
			"visualStyle": 1,
			"colorScheme": 0,
			"easing": 3,
			"toggleSize": 180
		}
	\end{verbatim}
	
	\textbf{Decoded indices:}
	\begin{verbatim}
		{
			"thumbShape_decoded": "circle",
			"visualStyle_decoded": "shadow",
			"colorScheme_decoded": "modern",
			"easing_decoded": "ease-out"
		}
	\end{verbatim}
	
\end{tcolorbox}

\end{document}